\documentclass[journal]{IEEEtran}
\usepackage{graphicx}
\usepackage{amsmath}
\usepackage{mathrsfs}
\usepackage{amsfonts}
\usepackage{algorithm, algcompatible}
\usepackage{color}
\usepackage{bm}
\usepackage{multirow}
\usepackage{amssymb}
\usepackage{mathdots}
\usepackage{enumerate}
\usepackage{setspace}
\usepackage{pifont}

\begin{document}

\title{Joint Reconfigurable Intelligent Surface Location and Passive Beamforming Optimization for Maximizing the Secrecy-Rate}

\author{Haiyan Guo, {\itshape{Member, IEEE}}, Zhen Yang,  {\itshape{Senior Member, IEEE}}, Yulong Zou$^*$, \itshape{Senior Member, IEEE},\\ 
\rm{Bin Lyu},  {\itshape{Member, IEEE}}, Yuhan Jiang, and Lajos Hanzo$^*$, \IEEEmembership{Fellow, IEEE}

\thanks{Copyright (c) 2015 IEEE. Personal use of this material is permitted. However, permission to use this material for any other purposes must be obtained from the IEEE by sending a request to pubs-permissions@ieee.org. }
\thanks{L. Hanzo would like to acknowledge the financial support of the Engineering and Physical Sciences Research Council projects EP/W016605/1 and EP/P003990/1 (COALESCE) as well as of the European Research Council's Advanced Fellow Grant QuantCom (Grant No. 789028). This work was supported in part by the National Natural Science Foundation of China under Grants 62071253, No.62071242, 62271268, No.61901229 and No. 62071005, in part by the Incubation Project of Nanjing University of Posts and Telecommunications for National Natural Science Foundation of China (No. NY220179).(Corresponding authors: Yulong Zou; Lajos Hanzo.)}
\thanks{H. Guo, Z. Yang, Y. Zou, B. Lyu and Y. Jiang are with the School of Communications and Information Engineering, Nanjing University of Posts and Telecommunications, Nanjing, China (E-mails: guohy@njupt.edu.cn; yangz@njupt.edu.cn; yulong.zou@njupt.edu.cn; blyu@njupt.edu.cn; yuhanjiang@njupt.edu.cn). }
\thanks{L. Hanzo is with the School of Electronics and Computer Science, University of Southampton, Southampton SO17 1BJ, U.K. (E-mail: lh@ecs.soton.ac.uk).}
}

\maketitle

\begin{abstract}
The physical layer security (PLS) is investigated for reconfigurable intelligent surface (RIS) assisted wireless networks,  where a source transmits its confidential information to a legitimate destination with the aid of a single small RIS in the presence of a malicious eavesdropper.  A new joint RIS location and passive beamforming (J-LPB) optimization scheme is proposed for the sake of maximizing the secrecy rate under the RIS location constraint and the constraint that the modulus of the reflecting coefficient at each RIS's unit is not larger than 1. Specifically, we analyze the optimal location of the RIS, and conclude that the product involving the source-RIS distance and the RIS-destination distance should be minimized. Since the product minimization problem is nonconvex, we then propose a two-tier optimization algorithm for solving it. Based on the near-optimal RIS 3D location obtained, we further formulate the passive beamforming optimization problem, and then propose to apply the Charnes-Cooper transformation along with the sequential rank-one constraint relaxation (SROCR) algorithm to solve it. Our numerical results show that the secrecy rate of the proposed J-LPB optimization scheme is higher than that of the benchmarks. Explicitly, we use the following benchmarks: the near-source-based RIS location and passive beamforming  (NSB-LPB)  optimization scheme,  the near-destination-based RIS location and passive beamforming  (NDB-LPB) optimization scheme, and the random RIS location and passive beamforming (R-LPB) optimization scheme. Finally, the benefits of our J-LPB scheme are further increased with the number of RIS units. 
\end{abstract}


\IEEEpeerreviewmaketitle

\section{Introduction}
\IEEEPARstart {W}{ireless} communication has become the norm. However, compared to wired channels, the wireless propagation environment is much more hostile, seriously degrade the wireless communication performance. In recent years, reconfigurable intelligent surfaces (RISs) have drawn extensive attention, since they are capable of intelligently reconfiguring the wireless propagation environment [1]-[7]. More explicitly, a RIS is a two-dimensional (2-D) metasurface consisting of a software based controller and a certain number of low-cost passive reflecting elements, each being able to reflect the incident signals with a tunable phase shift without dissipating any transmit power. In this way, it can reflect the incident electromagnetic waves with desirable phase shifts in a conveniently controllable way.   \\
\indent However, due to the broadcast nature of wireless links, the signals transmitted from the source to the legitimate users are always prone to malicious eavesdropping. Physical layer security (PLS), which guarantees confidential information transmission by exploiting the unique characteristics of wireless channels,  has been extensively explored [8]-[10]. A popular PLS technique is to harness the assistance of active relays to enhance the signals received at the legitimate destination [11]-[15]. However, this solution results in non-negligible power consumption and hardware cost due to the fact that the active relays need power-thirsty active components for forwarding signals. Hence, it is more attractive to harness RISs as passive relays for passive RIS-aided beamforming since only a moderate amount of energy is required by the RIS controller. \\
\indent Hence, some efforts have already been devoted to RIS-aided PLS. In [16]-[19], the RIS  phase shift coefficients were optimized for maximizing the secrecy rate. In [20], Chu \textit{et} \textit{al.} studied the joint optimization of the transmit power allocation and the RIS reflecting phase shifts, aiming for minimizing the transmit power subject to a specific secrecy rate constraint. In [21], Guan \textit{et} \textit{al.} proposed to jointly optimize the  RIS-aided beamforming and the transmit beamforming of the source, harnessing the extra aid of artificial noise (AN) for maximizing the achievable secrecy rate. In [22], Chen \textit{et} \textit{al.} proposed to maximize the minimum secrecy rate by jointly optimizing the beamformers at the base station and the reflecting coefficients at the RIS for a broadcast system supporting multiple legitimate receivers and multiple malicious eavesdroppers.  In [23], a robust and secure beamforming design method was proposed for maximizing the minimum achievable secrecy rate for the RIS-assisted millimeter wave communication. In [24],  the joint design of the transmit beamformers and of the AN at the access point as well as of the RIS phase shifters was proposed for maximizing the system's sum-rate, while limiting the achievable rate at each eavesdropper.  In [25], Tuan \textit{et} \textit{al.} proposed a tile allocation and phase shift adjustment strategy for maximizing the average secrecy rate (ASR) of RIS-aided indoor transmissions and evaluated the corresponding secrecy outage probability (SOP).  In [26], Tang \textit{et} \textit{al.} investigated the PLS of RIS-aided non-orthogonal multiple access (NOMA) networks and derived the closed-form expressions of both the SOP and of the ASR.  In [27], Hong \textit{et} \textit{al.} exploited a RIS to enhance the PLS of AN-aided multiple-input multiple-output systems, and jointly optimized the transmit precoding matrix and the AN covariance matrix as well as the RIS phase shifts for maximizing the system's secrecy rate. In [28], Kawai \textit{et} \textit{al.} proposed to jointly optimize the transmit beamforming weights and phase-shifting matrix at the RIS for minimizing the transmit power under the security constraints. In [29], Li \textit{et} \textit{al.} investigated the joint optimization of the beamforming matrix, AN matrix, phase shift matrix at the multiple RISs and surface selection matrix to guarantee the worst-case achievable system secrecy rate among multiple users. \\
\indent However, the RIS location has a substantial impact on the performance of RIS assisted wireless networks, as it directly affects the wireless channels involved. In [30], Kishk \textit{et} \textit{al.} studied how randomly equipping a fraction of blockages with RISs to improve the lower bound of the performance of the cellular network. In [31], Tao \textit{et} \textit{al.} derived a closed-form upper bound for the ergodic capacity of RIS-aided communication systems and showed that it is better to deploy the RIS near either the transmitter or the receiver for the sake of increasing the capacity. In [32],  Lu \textit{et} \textit{al.} proposed to jointly optimize the transmit beamforming of the source, the horizontal coordinate of a RIS in the air and the phase shifts for maximizing the worst-case signal-to-noise ratio (SNR).  In [33],  Makarfi \textit{et} \textit{al.} investigated the PLS of a vehicular network with the aid of RIS-based relaying and showed the effect of the RIS-relay location on the secrecy capacity. In [34], Hashida \textit{et} \textit{al.} proposed to optimize the RIS placement for maximizing the mean signal-to-interference-plus-noise (SINR) of the RIS-aided aerial-to-ground cellular networks. In [35], Zeng \textit{et} \textit{al.} analyzed the coverage of a RIS-aided downlink cellular network and optimized the RIS orientation as well as the horizontal distance for the sake of  maximizing the cell's coverage quality. \\
\indent As stated above, RIS-aided PLS can be enhanced by involving passive beamforming optimization. Moreover, the performance of RIS-aided wireless networks is critically affected by the RIS locations. Thus, in this paper we study the joint optimization of RIS location and passive beamforming for improving the PLS of a RIS-assisted wireless network operating without a direct source-destination link. Specifically, we consider a wireless network, where a source sends its confidential information to a legitimate destination in the presence of a malicious eavesdropper with the aid of a small RIS.  We propose a joint RIS location and passive beamforming optimization scheme for the sake of maximizing the secrecy rate of the RIS-aided wireless system. Our contributions are boldly and explicitly contrasted to the existing solutions in Table $\textrm{\uppercase\expandafter{\romannumeral1}}$, where there are three unique hitherto unsolved issues, which are summarized as follows.      
\begin{table*} 
\caption{ Contrasting our contributions to the current RIS-aided PLS techniques }
\small
 \begin{center}
  \begin{tabular}{|l|c|c|c|c|c|c|c|c|}
    \hline
Feature&\!\!Proposed\!\!\!&\!\!\!\!\!\! [17]-[23]\!\!\! \!\!&\!\!\![24]\!\!\! \! \!\!\!&\!\!\!\!\! [25] \!\!\!\!\! &\!\!\!\!\! [26]\!\!\!\! \!&\!\!\!\!\! [27]\!\!\!\! \!&\!\!\!\!\! [28]\!\!\!\! \! &\!\!\!\!\! [29]\!\!\!\! \!  \\ \hline
\!\!Secrecy rate maximization&\checkmark&\checkmark&\quad&\checkmark&\quad&\checkmark&\quad&\checkmark  \\\hline
\!\!\textbf{Joint RIS location and passive beamforming optimization}\!\!\!&\checkmark&\quad&\quad&\quad&\quad&\quad&\quad&\quad  \\\hline
\!\!\textbf{RIS location analysis and optimization}&\checkmark&\quad&\quad&\quad&\quad&\quad&\quad&\quad \\\hline
\!\!Phase shift optimization&\checkmark&\checkmark&\checkmark&\checkmark&\checkmark&\checkmark&\checkmark&\checkmark   \\\hline
\!\!\textbf{Amplitude reflection coefficient optimization}&\checkmark&\quad&\quad&\quad& \quad&\quad&\quad&\quad  \\\hline
 \end{tabular}
 \end{center}
\end{table*}
\begin{enumerate} 
\item We formulate the joint RIS location and passive beamforming optimization problem, which aims to maximize the secrecy rate of the system under the RIS location constraint and the constraint that the modulus of the reflecting coefficient at each RIS's unit is not larger than 1. By analyzing the expression of the optimization function, we  derive the conclusion that the 3D RIS coordinate should be optimized for the sake of minimizing a specific product, which involves both the source-RIS distance and the RIS-destination distance. Since the product minimization problem is non-convex, we propose a two-tier optimization algorithm for solving it to obtain a near-optimal 3D RIS coordinate. 
\item Based on the near-optimal RIS location obtained, we further formulate the passive beamforming optimization problem, and propose a  sequential rank-one constraint relaxation (SROCR) based iterative algorithm to solve it. Specifically, we first transform the passive beamforming optimization problem to a rank-one semidefinite matrix optimization problem by applying the Charnes-Cooper transformation. Then, we use the SROCR algorithm  to solve the semidefinite matrix optimization problem by partially relaxing the non-convex rank-one constraint in an iterative manner. 
\item Our numerical results show that the proposed joint RIS location and passive beamforming (J-LPB) optimization scheme outperforms the benchmarks in terms of its secrecy rate. Explicitly, we use the near-source-based RIS location and passive beamforming (NSB-LPB) optimization scheme, the near-destination-based RIS location and passive beamforming (NDB-LPB) optimization scheme and finally the random location and passive beamforming (R-LPB) optimization scheme. Additionally, the secrecy rate of the proposed J-LPB optimization scheme further increases with the number of reflecting elements.  
\end{enumerate}
\par\indent The outline of the paper is as follows. Section $\textrm{\uppercase\expandafter{\romannumeral2}}$ introduces the system model and presents our secrecy rate analysis. In Section $\textrm{\uppercase\expandafter{\romannumeral3}}$, the joint optimization of RIS location and passive beamforming is investigated.  Our simulation results are provided in Section $\textrm{\uppercase\expandafter{\romannumeral4}}$, while Section $\textrm{\uppercase\expandafter{\romannumeral5}}$ concludes the paper.\\
\indent \textit{Notation}: Bold letters are used for the vectors or matrices; $(\cdot)^{\rm{T}}$ represents the transpose and $(\cdot)^{\rm{H}}$ denotes the conjugate; $|\cdot|$ is the absolute value of a complex scalar and $||\cdot||$ is the $l_2$ norm of a vector; $\lambda_{max}(\cdot)$ denotes the largest eigenvalue of a matrix and $\bm{u}_{max}(\cdot)$ denotes the eigenvector associated with $\lambda_{max}(\cdot)$; $tr(\cdot)$ denotes the trace of a matrix and $rank(\cdot)$ denotes the rank of a matrix; $diag(\cdot)$ denotes a diagonal matrix whose diagonal elements are the corresponding elements of a vector; $\lceil  \cdot \rceil$ represents the rounding up operation and $\lfloor \cdot \rfloor$ represents the rounding down operation. 
\section{System Model and Secrecy Rate Analysis}
As shown in Fig. 1, we consider an indoor wireless network consisting of a source $S$, a small RIS $I$, a legitimate destination $D$ and a malicious eavesdropper $E$. In the system, the direct link from $S$ to $D$ is assumed to be blocked by a bulky obstacle. A small RIS having $N$ reflecting units is employed as a passive relay to assist transmission from $S$ to $D$. Explicitly, $S$ communicates with $D$ via the RIS. Furthermore, $E$ is located in the vicinity of $D$, pretending to be a legitimate receiver as well as to guarantee having an $I$-$E$ link in the presence of an $I$-$D$ link for wiretapping. This may happen in a scenario, where a legitimate user near the destination $D$ is captured by a Trojan and slaved as an eavesdropper. Following [24], we assume that $S$ focuses its transmit beams on the RIS. Due to the fact that $E$ is near $D$, we can assume that $E$ cannot intercept the link between $S$ and $I$. In the wireless network,  $S$, $D$ and $E$ are single-antenna nodes. It is noted that the small RIS does not have to be on the wall. It can be also fixed on a board  which is easily to be placed. In this way, the RIS can be placed more flexibly in a larger place. \\  
\begin{figure}[htb!]
\centering
\includegraphics[scale=1]{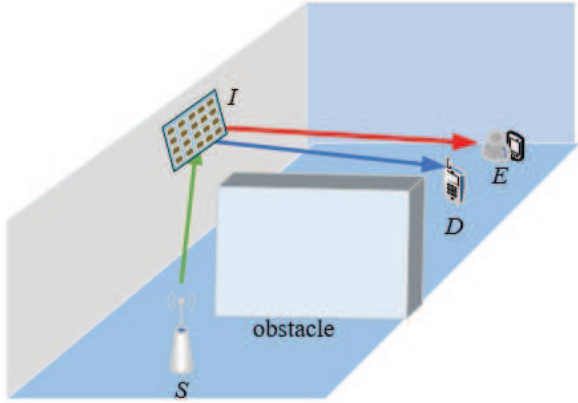}
\caption{ An indoor wireless network consisting of a source, a small RIS, a legitimate destination and an eavesdropper. }
\label{fig_1}
\end{figure}
\indent Let us denote the 3D coordinates of $S$, $D$, $E$ and the center of $I$ as $\bm{p}_S=\lbrack x_S, y_S, z_S \rbrack$, $\bm{p}_D=\lbrack x_D, y_D, z_D \rbrack$, $\bm{p}_E=\lbrack x_E, y_E, z_E \rbrack$ and $\bm{p}_I=\lbrack x_I, y_I, z_I \rbrack$, respectively.  Due to the fact that both the reflecting units and the gaps between the units are very small [36] [37], we can approximately denote the coordinate of each unit of the RIS as $\bm{p}_I$ in the case where the RIS is small. Then, we can obtain the distance between $S$ and $I$, that between $I$ and $D$, and that between $I$ and $E$ as $d_{SI}=||\bm{p}_I-\bm{p}_S||$, $d_{ID}=||\bm{p}_I-\bm{p}_D||$ and $d_{IE}=||\bm{p}_I-\bm{p}_E||$, respectively. \\
\indent In this paper, we consider the case where the RIS is deployed in the far field of $S$, $D$ and $E$. That is, we have $d_{SI} \geq d_{f}$, $d_{ID} \geq d_{f}$ and $d_{IE} \geq d_{f}$, where $d_{f}$ denotes the boundary of the far field and the near field. Upon denoting the path loss exponent of the $S$-$I$ link, the $I$-$D$ link and the $I$-$E$ link as $\alpha_{SI}$, $\alpha_{ID}$  and $\alpha_{IE}$, respectively, we can obtain the channel coefficients from $S$ to $I$, from $I$ to $D$ and from $I$ to $E$ as
\begin{equation}\label{(1)}
\bm{h}_{SI}=\sqrt{ \frac {\xi_0}{d_{SI}^{\alpha_{SI}} } } \widetilde{{\bm{h}}}_{SI},
\end{equation}
\begin{equation}\label{(2)}
\bm{h}_{ID}=\sqrt{ \frac {\xi_0}{d_{ID}^{\alpha_{ID}} } } \widetilde{{\bm{h}}}_{ID}
\end{equation}
and
\begin{equation}\label{(3)}
\bm{h}_{IE}=\sqrt{ \frac {\xi_0}{d_{IE}^{\alpha_{IE}} } } \widetilde{{\bm{h}}}_{IE},
\end{equation}
respectively, where $\xi_0$ is the path loss value at the reference distance of $d_0=1$ m, and $\widetilde{{\bm{h}}}_{SI}  \in \mathbb{C}^{N\times 1}$, $\widetilde{{\bm{h}}}_{ID}  \in \mathbb{C}^{N\times 1}$ and  $\widetilde{{\bm{h}}}_{IE}  \in \mathbb{C}^{N\times 1}$  are the small fading coefficients of the $S$-$I$ channel, the $I$-$D$ channel and the $I$-$E$ channel. \\
\indent The source $S$ transmits a signal $s$ to $D$, which has zero mean and unit variance, given the transmission power $P_s$ with the aid of the RIS $I$. Upon introducing  $\beta_n \in \lbrack 0,1 \rbrack$ and $\theta_n \in \lbrack 0,2 \pi \rbrack$, $n=1,2,...,N$, as the amplitude reflection coefficient and phase shift of the  $n$th reflecting unit of the RIS, respectively, we can model the reflection by the $n$th unit as $q_n=\beta_n e^{j\theta_n}$ and express the reflection by the RIS as $\bm{q}=[q_1,q_2,...,q_N]$. \\
\indent {Let us denote the small scale fading coefficients of the cascaded $S-I-D$ channel and the cascaded $S-I-E$ channel as $\widetilde{{\bm{h}}}_{SID}=diag(\widetilde{{\bm{h}}}_{SI}) \widetilde{{\bm{h}}}_{ID}$ and $\widetilde{{\bm{h}}}_{SIE}=diag(\widetilde{{\bm{h}}}_{SI}) \widetilde{{\bm{h}}}_{IE}$, respectively. Then, the instantaneous signal-to-noise ratio (SNR) at $D$  and that at $E$ are given by
\begin{equation}\label{(4)}
\gamma_{d}(\bm{q},\bm{p}_I)=\frac{P_s \xi_0^2 | \bm{q} {\widetilde{{\bm{h}}}_{SID}}|^2}{d_{ID}^{ \alpha_{ID}} d_{SI}^{\alpha_{SI}} \delta_n^2 }
\end{equation}
and
\begin{equation}\label{(5)}
\gamma_{e}(\bm{q},\bm{p}_I)=\frac{P_s \xi_0^2 | \bm{q} {\widetilde{{\bm{h}}}_{SIE}}|^2}{d_{IE}^{ \alpha_{IE}} d_{SI}^{\alpha_{SI}} \delta_n^2 },
\end{equation}
respectively, where $\delta_{n}^2$ is the variance of the additive white Gaussian noise (AWGN) received at $D$ and $E$. Thus, we can express the achievable secrecy rate as 
\begin{align}\label{(6)}
C_s&={\mathrm{max}}(0, \log_2 \lbrack 1+\gamma_{d}(\bm{q},\bm{p}_I) \rbrack-\log_2 \lbrack 1+\gamma_{e}(\bm{q},\bm{p}_I) \rbrack) \\ \nonumber
&=\!{\mathrm{max}} (0, \log_2 \! \frac{ d_{ID}^{\alpha_{ID}} d_{SI}^{\alpha_{SI}}  d_{IE}^{\alpha_{IE}} \delta_n^2 \!+\!\! P_s \xi_0^2  d_{IE}^{\alpha_{IE}} | \bm{q} {\widetilde{{\bm{h}}}_{SID}}|^2 }{ d_{ID}^{\alpha_{ID}} d_{SI}^{\alpha_{SI}}  d_{IE}^{\alpha_{IE}} \delta_n^2 \!+\!\!P_s \xi_0^2 d_{ID}^{\alpha_{ID}} | \bm{q} {\widetilde{{\bm{h}}}_{SIE}}|^2 }  \! ).
\end{align}
\indent Let us now introduce the notation of $f_1(\bm{p}_I)=d_{ID}^{\alpha_{ID}} d_{SI}^{\alpha_{SI}} d_{IE}^{\alpha_{IE}}$, $f_2(\bm{p}_I)=P_s \xi_0^2 d_{IE}^{\alpha_{IE}}$, and $f_3(\bm{p}_I)=P_s \xi_0^2 d_{ID}^{\alpha_{ID}}$. Then, we can rewrite (6) equivalently as
\begin{equation}\label{(7)}
C_s={\mathrm{max}} (0,\log_2  \frac{ \delta_n^2 f_1(\bm{p}_I ) + f_2(\bm{p}_I )  | \bm{q} {\widetilde{{\bm{h}}}_{SID}}|^2}{\delta_n^2 f_1(\bm{p}_I) +f_3(\bm{p}_I) | \bm{q} {\widetilde{{\bm{h}}}_{SIE}}|^2 }  ).
\end{equation}
\section{Joint Optimization of RIS location and Passive Beamforming}
\indent In this section, we jointly optimize the RIS location and the passive beamforming matrix for the sake of maximizing the secrecy rate of the system. We first formulate the joint optimization problem, and then analyze the optimal RIS location to conclude that  $d_{SI}^{\alpha_{SI}}d_{ID}^{\alpha_{ID}}$ should be minimized. After that, we propose a two-tier optimization algorithm to solve the $d_{SI}^{\alpha_{SI}}d_{ID}^{\alpha_{ID}}$ minimization problem to obtain a near-optimal RIS location. Then, based on the near-optimal RIS location obtained, we study the RIS beamforming optimization for the sake of maximizing the secrecy rate. Finally, we extend our work to a multiple-input single-output (MISO) system.\\
\indent Let us now aim for maximizing the system's secrecy rate by jointly optimizing the RIS' location $\bm{p}_I$ and passive  beamforming vector $\bm{q}$. From (7), the joint optimization problem is formulated as
\begin{eqnarray}\label{(8)} 
 [\bm{p}_I^{opt}, \bm{q}^{opt}]  &\!\!\!\!\!\!\!\!=\!\!\!\!\!\!\!\!&\mathrm{arg} \!\!\quad\!\! \underset{\bm{p}_I, \bm{q}}{\mathrm{max}} F(\bm{p}_I, \bm{q})\!\!=\!\!\frac{\delta_n^2 f_1(\bm{p}_I) \!\!+\!\! f_2(\bm{p}_I)  | \bm{q} {\widetilde{{\bm{h}}}_{SID}}|^2 }{\delta_n^2 f_1(\bm{p}_I)\!\!+\!\!f_3(\bm{p}_I) | \bm{q} {\widetilde{{\bm{h}}}_{SIE}}|^2 }  \nonumber\\
&& {}
\begin{array}{ll}
\mathrm{s.t.}  
&\bm{p}_I \in \Omega_p\\
&0 \leq |q_n| \leq 1,n=1,2,...,N,\\
\end{array}
\end{eqnarray}
where $\Omega_p$ denotes the specific geographic location set where the RIS should be placed in order to guarantee that neither the $S-I$ link nor the $I-D$ link is blocked by the obstacle. It is worth noting that the optimal RIS location should be fixed in all the transmission slots, while its passive beamforming vector can be optimized according to the instantaneous channel state information (CSI) in each transmission slot for improving the secrecy performance. 
\subsection{The optimization of 3D RIS location} 
\indent By substituting  $f_1(\bm{p}_I)=d_{ID}^{\alpha_{ID}} d_{SI}^{\alpha_{SI}} d_{IE}^{\alpha_{IE}}$, $f_2(\bm{p}_I)=P_s \xi_0^2 d_{IE}^{\alpha_{IE}}$ and $f_3(\bm{p}_I)=P_s \xi_0^2 d_{ID}^{\alpha_{ID}}$ into $F(\bm{p}_I, \bm{q})$, after a few steps, we can equivalently express $F(\bm{p}_I, \bm{q})$ as
\begin{align}\label{(9)}
F(\bm{p}_I, \bm{q})&=1+\frac{P_s \xi_0^2}{\delta_n^2} \frac{ | \bm{q} {\widetilde{{\bm{h}}}_{SID}}|^2-\frac{ d_{ID}^{\alpha_{ID}} } {d_{IE}^{\alpha_{IE}} }  | \bm{q} {\widetilde{{\bm{h}}}_{SIE}}|^2  }{ d_{SI}^{\alpha_{SI}} d_{ID}^{\alpha_{ID}}+ \frac{P_s \xi_0^2}{\delta_n^2}\frac{ d_{ID}^{\alpha_{ID}} } {d_{IE}^{\alpha_{IE}} }  | \bm{q} {\widetilde{{\bm{h}}}_{SIE}}|^2    } \\ \nonumber
&=1+\frac{P_s \xi_0^2}{\delta_n^2} 
\frac{ \bm{q}  ({\widetilde{{\bm{h}}}_{SID}} {\widetilde{{\bm{h}}}_{SID}}^{\rm{H}}\!\!-\!\!\frac{ d_{ID}^{\alpha_{ID}} } {d_{IE}^{\alpha_{IE}} } {\widetilde{{\bm{h}}}_{SIE}} {\widetilde{{\bm{h}}}_{SIE}}^{\rm{H}})  {\bm{q}}^{\rm{H}}  }
{ d_{SI}^{\alpha_{SI}} d_{ID}^{\alpha_{ID}}+ \frac{P_s \xi_0^2}{\delta_n^2}\frac{ d_{ID}^{\alpha_{ID}} } {d_{IE}^{\alpha_{IE}} }  | \bm{q} {\widetilde{{\bm{h}}}_{SIE}}|^2    } .
\end{align}
\indent Let us introduce the short-hand of $\bm{B}={\widetilde{{\bm{h}}}_{SID}} {\widetilde{{\bm{h}}}_{SID}}^{\rm{H}}-\frac{ d_{ID}^{\alpha_{ID}} } {d_{IE}^{\alpha_{IE}} } {\widetilde{{\bm{h}}}_{SIE}} {\widetilde{{\bm{h}}}_{SIE}}^{\rm{H}}$. We can then theoretically prove that $\bm{B}$ is NOT a negative semidefinite  matrix. The proof details are presented in the Appendix. Thus, we can obtain that there always exist a set of $\bm{q}$ values, which can guarantee   $ \bm{q} \bm{B} \bm{q}^{\rm{H}} \geq 0$. \\
\indent Since $\bm{q}^{opt}$ aims for maximizing $F(\bm{p}_I, \bm{q})$, $\bm{q}^{opt}$ should be selected from the set of  of $\bm{q}$ values, which satisfy $\bm{q} \bm{B} \bm{q}^{\rm{H}} \geq 0$. Thus, from (9), we readily see that  $d_{SI}^{\alpha_{SI}} d_{ID}^{\alpha_{ID}}$ should be minimized for the sake of maximizing $F(\bm{p}_I, \bm{q})$. Additionally, it is noted that  the value of  $\frac{ d_{ID}^{\alpha_{ID}} } {d_{IE}^{\alpha_{IE}} }$ is near-constant, regardless of the RIS location in our scenario, where $E$ is located in the vicinity of $D$. This means that the effect of the RIS location on $\frac{ d_{ID}^{\alpha_{ID}} } {d_{IE}^{\alpha_{IE}} }$ can be ignored.\\
\indent {As analyzed above, the RIS 3D location $\bm{p}_I$ should be optimized  for the sake of minimizing $d_{SI}^{\alpha_{SI}} d_{ID}^{\alpha_{ID}}$, regardless of the optimal RIS passive beamforming vector. Due to that $d_{SI}^{\alpha_{SI}} d_{ID}^{\alpha_{ID}}$ is independent of the RIS passive beamforming vector $\bm{q}$, we can first obtain the optimal RIS location by solving the $d_{SI}^{\alpha_{SI}} d_{ID}^{\alpha_{ID}}$ minimization problem separately. Moreover, in our scenario, once the RIS is deployed, its position is always fixed. Thus, we can then optimize the RIS passive beamforming vector $\bm{q}$ separately based on the obtained optimal RIS location. That is, the joint optimization problem (8) can be decoupled into the RIS location optimization problem and the RIS passive beamforming optimization problem successively. \\
\indent Now, we optimize the 3D RIS coordinate for the sake of minimizing $d_{SI}^{\alpha_{SI}} d_{ID}^{\alpha_{ID}}$. We denote the coordinate of the lower left endpoint of the obstacle $O^{l,l}$ by $\bm{p}_o^{l,l}=[x_{o}^{l,l},y_{o}^{l,l},z_{o}^{l,l}]$, the coordinate of the upper left endpoint of the obstacle $O^{u,l}$ by $\bm{p}_o^{u,l}=[x_{o}^{u,l},y_{o}^{u,l},z_{o}^{u,l}]$, the coordinate of the lower right endpoint of the obstacle $O^{l,r}$ by $\bm{p}_o^{l,r}=[x_{o}^{l,r},y_{o}^{l,r},z_{o}^{l,r}]$ and the coordinate of the upper right endpoint of the obstacle $O^{u,r}$ by $\bm{p}_o^{u,r}=[x_{o}^{u,r},y_{o}^{u,r},z_{o}^{u,r}]$. For simplicity, here we do not consider the thickness of the obstacle, and focus our attention on the case, where both the lower right endpoint and the upper right endpoint of the obstacle whose height is $L_z$ are located at the wall. That is, $y_{o}^{l,r}=y_{o}^{u,r}=L_y$, $z_{o}^{u,l}=z_{o}^{u,r}=L_z$. It has to be pointed out that the proposed joint optimization of the RIS location and passive beamforming vector can be extended to a general case, where the square obstacle is located randomly between $S$ and $D$. \\ 
\indent Let us define $\beta_{11}=(y_{o}^{l,l}-y_S)(z_{o}^{u,l}-z_S)-(y_{o}^{u,l}-y_S)(z_{o}^{l,l}-z_S)$, $\beta_{12}=(x_{o}^{l,l}-x_S)(z_{o}^{u,l}-z_S)-(x_{o}^{u,l}-x_S)(z_{o}^{l,l}-z_S)$, $\beta_{13}=(x_{o}^{l,l}-x_S)(y_{o}^{u,l}-y_S)-(x_{o}^{u,l}-x_S)(y_{o}^{l,l}-y_S)$, $\beta_{21}=(y_{o}^{l,l}-y_D)(z_{o}^{u,l}-z_D)-(y_{o}^{u,l}-y_D)(z_{o}^{l,l}-z_D)$, $\beta_{22}=(x_{o}^{l,l}-x_D)(z_{o}^{u,l}-z_D)-(x_{o}^{u,l}-x_D)(z_{o}^{l,l}-z_D)$ and $\beta_{23}=(x_{o}^{l,l}-x_D)(y_{o}^{u,l}-y_D)-(x_{o}^{u,l}-x_D)(y_{o}^{l,l}-y_D)$. Upon denoting $\bm{\beta}_1=[\beta_{11}, -\beta_{12}, \beta_{13}]$, $\bm{\beta}_2=[\beta_{21}, -\beta_{22}, \beta_{23}]$, we can obtain the equation of the plane consisting of $S$, $O^{l,l}$ and $O^{u,l}$ and that consisting of $D$, $O^{l,l}$ and $O^{u,l}$ as
\begin{equation}\label{(10)}
\bm{\beta}_1 (\bm{p}-\bm{p}_S)^{\rm{T}}=0
\end{equation}
and
\begin{equation}\label{(11)}
\bm{\beta}_2 (\bm{p}-\bm{p}_D)^{\rm{T}}=0,  
\end{equation}
respectively, where $\bm{p}$ is the 3-D coordinate of the plot on the corresponding plane.  \\
\indent Let us now study the minimization of $d_{SI}^{\alpha_{SI}} d_{ID}^{\alpha_{ID}}$ by optimizing the 3D RIS coordinate. Let us denote $\bm{p}_I(1)$, $\bm{p}_I(2)$ and $\bm{p}_I(3)$ as the x-coordinate, y-coordinate and z-coordinate of $I$, respectively, $[x_{min}, x_{max}]$ as the x-coordinate range of $I$, $[y_{min}, y_{max}]$ as the y-coordinate range of $I$,  $[z_{min}, z_{max}]$ as the z-coordinate range of $I$. By substituting the expressions of  $d_{SI}=||\bm{p}_I-\bm{p}_S||$ and $d_{ID}=||\bm{p}_I-\bm{p}_D||$ into $d_{SI}^{\alpha_{SI}} d_{ID}^{\alpha_{ID}}$, we formulate the optimization problem as   
\begin{eqnarray}\label{(12)}
 \bm{p}_I^{opt}&=&\mathrm{arg} \quad \underset{\bm{p}_I}{\mathrm{min}} \quad g( \bm{p}_I)\!\!=\!\! ||\bm{p}_I- \bm{p}_S||^{\alpha_{SI}}||\bm{p}_I- \bm{p}_D||^{\alpha_{ID}} \nonumber\\
&& {}
\begin{array}{ll}
\mathrm{s.t.}  &C1:\quad \bm{p}_I(1) \in \lbrack x_{min}, x_{max} \rbrack\\
&C2:\quad \bm{p}_I(2) \in \lbrack y_{min}, y_{max} \rbrack \\
&C3:\quad \bm{p}_I(3) \in \lbrack z_{min}, z_{max} \rbrack \\
&C4:\quad ||\bm{p}_I- \bm{p}_S|| \geq d_{f} \\
&C5:\quad ||\bm{p}_I- \bm{p}_D|| \geq d_{f} \\
&C6:\quad ||\bm{p}_I- \bm{p}_E|| \geq d_{f} \\
&C7:\quad \bm{\beta}_1 (\bm{p}-\bm{p}_S)^{\rm{T}}>0 \\
&C8:\quad \bm{\beta}_2 (\bm{p}-\bm{p}_D)^{\rm{T}}<0 
\end{array}
\end{eqnarray}
where the constraints C4, C5 and C6 guarantee that $I$ is in the far field of $S$, $D$ and $E$,  the constraint C7 guarantees that the $S$-$I$ link is not blocked by the obstacle, and the constraint C8 guarantees that the $I$-$D$ link is not blocked by the obstacle.  \\
\indent Since Problem  (12) is non-convex, we propose a two-tier optimization algorithm to obtain its near-optimal solution. In the inner tier, we fix the z-coordinate of $I$, $\bm{p}^j_I(3)$, and optimize $\bm{p}^j_I$ for minimizing $g(\bm{p}_I)$. Upon introducing the notation of $\bm{p}_1=[x_{min},y_{min},z_{min}]$, $\bm{p}_2=[x_{min},y_{min},z_{max}]$, $\bm{p}_3=[x_{min},y_{max},z_{min}]$, $\bm{p}_4=[x_{min},y_{max},z_{max}]$, $\bm{p}_5=[x_{max},y_{min},z_{min}]$, $\bm{p}_6=[x_{max},y_{min},z_{max}]$, $\bm{p}_7=[x_{max},y_{max},z_{min}]$ and $\bm{p}_8=[x_{max},y_{max},z_{max}]$, we have $d_{SI,max}=\underset{k=1,...,8} {\mathrm{max}}\quad ||\bm{p}_S-\bm{p}_k||$ and $d_{ID,max}=\underset{k=1,...,8} {\mathrm{max}}\quad ||\bm{p}_D-\bm{p}_k||$. Then, given $\bm{p}_I^j(3)$, the inner-tier optimization problem can be formulated as Problem (13) stated on the following page. \\
\newcounter{TempEqCnt12}
\setcounter{TempEqCnt12}{12}
\setcounter{equation}{12}
\begin{figure*}
\begin{eqnarray}\label{(13)}
(\bm{p}_I^j)^{opt}&=&\mathrm{arg} \quad  \underset{\bm{p}_I^j}{\mathrm{min}} \quad g( \bm{p}_I^j)= ||\bm{p}_I^j- \bm{p}_S||^{\alpha_{SI}}||\bm{p}_I^j-\bm{p}_D||^{\alpha_{ID}} \nonumber\\
&& {}
\begin{array}{ll}
\mathrm{s.t.}  &C1,C2,C6,C7,C8\\
&C9:\quad d_{SI} \in \lbrack d_{f}, d_{SI,max} \rbrack \\
&C10:\quad \!\!d_{ID}\in \lbrack {\mathrm{max}} (|d_{SD}\!\!-\!d_{SI}|,d_f), d_{ID,max} \rbrack \\
&C11:\quad ||\bm{p}_I^j- \bm{p}_S||=d_{SI}\\
&C12:\quad ||\bm{p}_I^j- \bm{p}_D||= d_{ID}. 
\end{array}
\end{eqnarray}
\noindent\rule[-10pt]{\textwidth}{0.5pt}
\end{figure*}
\indent From C11 and C12 of Problem (13), we can readily obtain the potential feasible solution $(\bm{p}_I^j)^{pf}$ given $d_{SI} \in \lbrack d_{f}, d_{SI,max} \rbrack $ and $d_{ID} \in \lbrack {\mathrm{max}} (|d_{SD}-d_{SI}|,d_f), d_{ID,max} \rbrack$.  If $(\bm{p}_I^j)^{pf}$ also satisfies C1, C2, C6, C7 and C8, we can state that it is a true feasible solution denoted as $(\bm{p}_I^j)^f$ and then compute the corresponding $g[(\bm{p}_I^j)^f]$. By performing comparisons and updating the operations among all the $\lbrack d_{SI}, d_{ID} \rbrack$ cases, we can obtain the minimal $g(\bm{p}_I^j)$ and the corresponding $(\bm{p}_I^j)^{opt}$ given $\bm{p}_I^j(3)$. It is worth noting that we solve Problem (13) by the search over the $d_{SI}$-axis and the $d_{ID}$-axis, not over the $x_I$-axis and the $y_I$-axis as in the conventional search method. The reason for this is that, since Problem (12) aims for minimizing $d_{SI}^{\alpha_{SI}} d_{ID}^{\alpha_{ID}}$,  it is more likely to obtain a more uniform and finer search grid for $d_{SI}^{\alpha_{SI}} d_{ID}^{\alpha_{ID}}$ by the search over the $d_{SI}$-axis and the $d_{ID}$-axis.\\
\indent In the outer-tier, we compare all the minimal $g(\bm{p}_I^j)$ to obtain the near-optimal solution  $\bm{p}_I^{opt}$ of Problem (12). That is,
\begin{eqnarray}\label{(14)}
\bm{p}_I^{opt}&\!\!\!=\!\!\!&\mathrm{arg} \quad \underset{j}{\mathrm{min}} \quad g(\bm{p}_I^j) \nonumber\\
&& {}
\begin{array}{ll}
\mathrm{s.t.}  &C3.
\end{array}
\end{eqnarray}
The details of the two-tier optimization algorithm are presented in Table $\text{\uppercase\expandafter{\romannumeral2}}$. \\
\begin{table}[!h]
\caption{Two-tier Optimization Algorithm}
\vspace{-0.7cm}
\begin{center}
\begin{tabular}{l}
\hline
Algorithm 1: Two-tier Optimization Algorithm.\\
\hline
\textbf{Function}
\emph{Outer{\_} Iteration}\\
\small
Step 1: Set the search step length $\delta_z$.\\
\small
Step 2: Initialize $j=0$, $z_I^j=0$,  $\bm{p}_I=[0,0,z_I^j]$, \\
\small
\quad \quad \quad \quad \quad \quad \quad $g(\bm{p}_I^{opt})\!\!=\!\!10^{20}$.\\
\small
Step 3: \textbf{While} $z_I^j \leq z_{max}$\\
\small
Step 4: Call \textbf{Function} \emph{Inner{\_}Optimization}. \\
\small
Step 5: \textbf{If} $g(\bm{p}_I^j) < g(\bm{p}_I^{opt})$ \\
\small
Step 6: Update \\ 
\small
\quad \quad \quad \quad \quad $g(\bm{p}_I^{opt})=g(\bm{p}_I^j)$ ,\\
\small
\quad \quad \quad \quad \quad $\bm{p}_I^{opt}=\bm{p}_I^j$.\\
\small
Step 7: \textbf{end If}\\
\small
Step 8: Increase $j=j+1$ and $z_I^j=z_I^j+\delta_z$.\\
\small
Step 9: \textbf{end While}\\
\small
Step 10: \textbf{Return} $\bm{p}_I^{opt}$.\\
\small
Step 11: Obtain the near-optimal solution $\bm{p}_I^{opt}$ of Problem (12). \\
\small
\textbf{end} \\
\small
\textbf{Function}
\emph{Inner{\_}Optimization} \\
\small
Step 12:  Set search step $\delta_d$. \\
\small
Step 13:  Initialize $d_{SI}=d_f $, $d_{ID}={\mathrm{max}} (|d_{SD}-d_{SI}|, d_f )$, \\
\small
\quad \quad \quad \quad \quad \quad \quad \quad  $\bm{p}_I^j=[0,0,z_I^j]$ and $g(\bm{p}_I^j)=10^{20}$. \\
\small
Step 14: \textbf{While} $d_{SI} \leq d_{SI,max}$ \\
\small
Step 15: \textbf{While} $d_{ID} \leq d_{ID,max}$ \\
\small
Step 16: Apply the ``solve" tool of Matlab to find the solution  \\
\small
\quad \quad \quad \quad $[x_I^j, y_I^j]$ to equations C11 and C12 in Problem (13),  \\
\small
\quad \quad \quad \quad and obtain a potential feasible solution $(\bm{p}_I^{j})^{pf}\!\!=\!\![x_I^j, y_I^j, z_I^j]$.   \\
\small
\quad \quad \quad \quad If $[x_I^j, y_I^j]$ is empty, set $(\bm{p}_I^{j})^{f}\!\!=\!\![0, 0, 0]$ and $g[(\bm{p}_I^{j})^{f}]\!\!=\!\!10^{20}$.\\
\small
Step 17: \textbf{If} $(\bm{p}_I^{j})^{pf}$ satisfies C1, C2, C6, C7  and C8 in Problem (13) \\
\small
Step 18:  Set $(\bm{p}_I^{j})^{f}=(\bm{p}_I^{j})^{pf}$ and compute $g[(\bm{p}_I^{j})^{f}]$. \\
\small
Step 19: \textbf{else}, set $(\bm{p}_I^{j})^{f}=[0, 0, 0]$ and $g[(\bm{p}_I^{j})^{f}]=10^{20}$. \\
\small
Step 20: \textbf{end if}\\
\small
Step 21: \textbf{If} $g[(\bm{p}_I^{j})^{f}] < g(\bm{p}_I^{j})$ \\
\small
Step 22: Update\\
\small
\quad \quad \quad \quad \quad \quad \quad $ g(\bm{p}_I^{j})=g[(\bm{p}_I^{j})^{f}]$, \\
\small
 \quad \quad \quad \quad \quad \quad \quad $\bm{p}_I^{j}=(\bm{p}_I^{j})^{f}$. \\
\small
Step 23: \textbf{end If}\\
\small
Step 24: Increase $d_{ID}=d_{ID}+\delta_d$.\\
\small
Step 25: \textbf{end While}\\
\small
Step 26: Increase $d_{SI}=d_{SI}+\delta_d$.\\
\small
Step 27: \textbf{end While}\\
\small
Step 28: \textbf{Return} $\bm{p}_I^j$ and $g(\bm{p}_I^j)$.   \\
\small
\textbf{end}\\
\hline
\end{tabular}
\end{center}
\end{table}
\indent Let us now discuss the computational complexity of the proposed two-tier optimization algorithm. Since $(\bm{p}_I^j)^{pf}$, $(\bm{p}_I^j)^f$, $\bm{p}_I^{j}$, $\bm{p}_I^{opt}$, $\bm{p}_S$ and $\bm{p}_D$  are three-dimensional, both the number of floating point operations $M_1$ in each inner-tier iteration, and that in each outer-tier iteration, namely $M_2$, are small. Let us introduce $d_{SI}=d_f+k\delta_d$, where $k$ is an integer in the range of 0 to $\frac{d_{SI,max}-d_f}{\delta_d}$ and analyze the computational complexity of the inner-tier optimization. \\
\indent When $k \in [0,\lfloor \frac{d_{SD}-2d_f}{\delta_d} \rfloor]$, from C10 in Problem (13), we have $d_{ID} \in [d_{SD}-d_{SI},d_{ID,max}]$. Then, we can obtain the number of floating point operations $K_1$ as
\begin{align}\label{(15)}
K_1&=\sum\limits_{k=0}^{\lfloor \frac{d_{SD}-2d_f}{\delta_d} \rfloor} \lfloor \frac{d_{ID,max}-(d_{SD}-d_f-k\delta_d)}{\delta_d} \rfloor M_1 \\ \nonumber
&\approx \frac{M_1}{\delta_d^2} (d_{SD}d_{ID,max}-\frac{1}{2}d_{SD}^2).
\end{align}
\indent When $k \in [\lceil \frac{d_{SD}-2d_f}{\delta_d} \rceil, \lfloor \frac{d_{SD}}{\delta_d} \rfloor]$, from C10 in Problem (13), we have $d_{ID} \in [d_f, d_{ID,max}]$. Then,  the number of floating point operations $K_2$ becomes:
\begin{equation}\label{(16)}
K_2=\sum\limits_{k=\lceil \frac{d_{SD}-2d_f}{\delta_d} \rceil}^{ \lfloor \frac{d_{SD}}{\delta_d} \rfloor} \lfloor \frac{d_{ID,max}- d_f}{\delta_d} \rfloor M_1 \approx 2 \frac{M_1}{\delta_d^2} d_f d_{ID,max}.
\end{equation}
\indent When $k \in [\lceil  \frac{d_{SD}}{\delta_d} \rceil, \lfloor \frac{d_{SI,max}-d_f}{\delta_d} \rfloor]$, from C10 in Problem (13), we have $d_{ID} \in [d_{SI}-d_{SD}, d_{ID,max}]$. Then, the corresponding number of floating point operations $K_3$ is given by (17) on the following page.
\newcounter{TempEqCnt100}
\setcounter{TempEqCnt100}{16}
\setcounter{equation}{16}
\begin{figure*}
\begin{equation}\label{(17)}
K_3=\sum\limits_{k=\lceil  \frac{d_{SD}}{\delta_d} \rceil}^{\lfloor \frac{d_{SI,max}-d_f}{\delta_d} \rfloor} \lfloor \frac{d_{ID,max}+(d_{SD}-d_f-k\delta_d)}{\delta_d} \rfloor M_1 \approx\frac{M_1}{\delta_d^2} (d_{ID,max}d_{SI,max}- d_{SD}d_{ID,max}+\frac{d_{SD}^2}{2}-\frac{d_{SI,max}^2}{2})
\end{equation}
\end{figure*}\\
\indent By combining (15), (16) and (17), we arrive at the total number of floating point operations of the inner-tier iteration as $K$ given by (18) on the following page. If we assume that $d_{SI,max}$ and $d_{ID,max}$ are of the same order of magnitude, then we have $K\approx \frac{M_1}{\delta_d^2} \frac{d_{ID,max}d_{SI,max}}{2}$. \\
\newcounter{TempEqCnt101}
\setcounter{TempEqCnt101}{17}
\setcounter{equation}{17}
\begin{figure*}
\begin{equation}\label{(18)}
K=K_1+K_2+K_3\approx\frac{M_1}{\delta_d^2} (d_{ID,max}d_{SI,max}-\frac{d_{SI,max}^2}{2}+2 d_f d_{ID,max})
\end{equation}
\end{figure*}
\indent By taking into account both the inner-tier and the outer-tier iterations, we can quantify the computational complexity of the proposed two-tier iterative algorithm approximately given as $\frac{z_{max}}{\delta_z} (\frac{M_1}{\delta_d^2} \frac{d_{ID,max}d_{SI,max}}{2}+M_2)$, which is lower than that of the traditional greedy search algorithm $\frac{z_{max}}{\delta_z} (\frac{M_1}{\delta_d^2} d_{ID,max}d_{SI,max}+M_2)$. 
\subsection{The RIS beamforming optimization}
\indent By substituting $\bm{p}_I^{opt}$ obtained by our proposed two-tier optimization algorithm into (8), we can determine $\bm{q}^{opt}$ by solving the following optimization problem  
\begin{eqnarray}\label{(19)}
\bm{q}^{opt} \!\!\!\!\!&=&\!\!\!\!\!\mathrm{arg} \quad\!\!\! \underset{\bm{q}}{\mathrm{max}}F(\bm{p}_I^{opt}, \bm{q})=\frac{\delta_n^2 f_1(\bm{p}_I^{opt}) \!\!+\!\! f_2(\bm{p}_I^{opt}) \bm{q} \widetilde{{\bm{H}}}_{SID}  \bm{q}^{\rm{H}}   }{\delta_n^2 f_1(\bm{p}_I^{opt})\!\!+\!\!f_3(\bm{p}_I^{opt}) \bm{q} \widetilde{{\bm{H}}}_{SIE}  \bm{q}^{\rm{H}}   }  \nonumber\\
&& {}
\begin{array}{ll}
\mathrm{s.t.}  
&0 \leq |q_n| \leq 1, n=1,2,...,N.\\
\end{array}
\end{eqnarray}        
\!where $\widetilde{{\bm{H}}}_{SID}= \widetilde{{\bm{h}}}_{SID} \widetilde{{\bm{h}}}_{SID}^{\rm{H}}$ and  $\widetilde{{\bm{H}}}_{SIE}= \widetilde{{\bm{h}}}_{SIE} \widetilde{{\bm{h}}}_{SIE}^{\rm{H}}$. In this paper, we assume that  the cascaded CSI can be obtained by using the existing CSI acquisition techniques, such as the joint bilinear factorization and matrix completion (JBF-MC) algorithm [38], the discrete Fourier transform (DFT) based method [39], the channel training-based protocol (CTBP) [40], the low-complexity arrangement partition based channel estimation protocol [41], etc. It is worth noting that the constraint $ |\bm{q} {\widetilde{{\bm{h}}}_{SID}}|^2-\frac{ d_{ID}^{\alpha_{ID}} } {d_{IE}^{\alpha_{IE}} }  | \bm{q} {\widetilde{{\bm{h}}}_{SIE}}|^2 \geq 0$ is ignored, since the solution which maximizes $F(\bm{p}_I^{opt}, \bm{q})$ always satisfies this constraint. The reason for this has been stated below (9).\\
\indent Upon introducing $\bm{\Psi}=\bm{q}^{\rm{H}} \bm{q}$, we can readily have $rank(\bm{\Psi})=1$ and $\bm{\Psi} \geq 0$. Then, we can obtain that $\bm{q} \widetilde{{\bm{H}}}_{SID}  \bm{q}^{\rm{H}}=tr(\widetilde{{\bm{H}}}_{SID} \bm{\Psi})$ and $\bm{q} \widetilde{{\bm{H}}}_{SIE}  \bm{q}^{\rm{H}}=tr(\widetilde{{\bm{H}}}_{SIE} \bm{\Psi})$.  Due to $0 \leq  |q_n| \leq 1 (n=1,2,...,N)$ and $\Psi_{n,n}=|q_n|^2$, where $\Psi_{n,n}$ denotes the element in the $n$th row and the $n$th column of $\bm{\Psi}$, we have $0 \leq \Psi_{n,n} \leq 1$. Thus, we can express Problem (19) as 
\begin{eqnarray}\label{(20)}
\bm{\Psi}^{opt} &\!\!\!=\!\!\!&\mathrm{arg} \quad \underset{\bm{\Psi}}{\mathrm{min}} \frac{\delta_n^2 f_1(\bm{p}_I^{opt})+f_3(\bm{p}_I^{opt}) tr(\widetilde{{\bm{H}}}_{SIE} \bm{\Psi})  }  {\delta_n^2 f_1(\bm{p}_I^{opt}) + f_2(\bm{p}_I^{opt}) tr(\widetilde{{\bm{H}}}_{SID} \bm{\Psi})   } \nonumber\\
&& {}
\begin{array}{ll}
\mathrm{s.t.}  
&\bm{\Psi} \geq 0\\
&0 \leq \Psi_{n,n} \leq 1, n=1,2,...,N\\
&rank (\bm{\Psi}) =1.\\
\end{array}
\end{eqnarray}
\indent Similar to [17], we apply the Charnes-Cooper transformation to cast Problem (20) into the equivalent non-fractional form. Specifically, upon introducing
\begin{equation}\label{(21)}
\bm{\Phi}=\frac{ \bm{\Psi} }{\delta_n^2 f_1(\bm{p}_I^{opt})+f_2(\bm{p}_I^{opt}) tr( \widetilde{{\bm{H}}}_{SID} \bm{\Psi}) }
\end{equation}
and
\begin{equation}\label{(22)}
t=\frac{1}{\delta_n^2 f_1(\bm{p}_I^{opt})+f_2(\bm{p}_I^{opt}) tr( \widetilde{{\bm{H}}}_{SID} \bm{\Psi}) },
\end{equation}
we can express (20)  as (23) stated on the following page. \\
\newcounter{TempEqCnt102}
\setcounter{TempEqCnt102}{22}
\setcounter{equation}{22}
\begin{figure*}
\begin{eqnarray}\label{(23)}
[\bm{\Phi}^{opt},t^{opt}] &=&\mathrm{arg}  \quad \underset{\bm{\Phi}, t}  {\mathrm{min}} \quad g(\bm{\Phi},t)=\delta_n^2 f_1(\bm{p}_I^{opt}) t + f_3(\bm{p}_I^{opt})  tr( \widetilde{{\bm{H}}}_{SIE} \bm{\Phi} )  \nonumber\\
&& {}
\begin{array}{ll}
\mathrm{s.t.}  
&\bm{\Phi} \geq 0\\
&0\leq \Phi_{n,n} \leq t, n=1,2,...,N\\
&\delta_n^2 f_1(\bm{p}_I^{opt}) t + f_2(\bm{p}_I^{opt})  tr( \widetilde{{\bm{H}}}_{SID} \bm{\Phi} )=1\\
& rank(\bm{\Phi})=1\\
\end{array}
\end{eqnarray}
\noindent\rule[-10pt]{\textwidth}{0.5pt}
\end{figure*}
\indent Observe from Problem (23) that the optimization function $g(\bm{\Phi},t)$ is convex and the constraint $rank(\bm{\Phi})=1$ is non-convex. To address this problem, we use the SROCR algorithm of [42], which gradually relaxes the non-convex rank-one constraint.  Specifically, we solve a convex  problem, where the rank-one constraint is partially relaxed in each iteration, and then update the solution iteratively. The details are presented as follows.\\
\indent Following [42], we first obtain $[\bm{\Phi}^{(0)},t^{(0)}]$ by solving the relaxed problem given by
\begin{eqnarray}\label{(24)}
[\bm{\Phi}^{(0)},t^{(0)}] &\!\!\!=\!\!\!&\mathrm{arg} \quad \underset{\bm{\Phi}\geq 0, t}{\mathrm{min}} g(\bm{\Phi},t)  \nonumber\\
&& {}
\begin{array}{ll}
\!\!\!\!\!\!\!\!\!\!\!\!\mathrm{s.t.}  \!\!\!\!\!\!
& \!\!\!\!\!0\leq \Phi_{n,n} \leq t, n=1,2,...,N\\
& \!\!\!\!\!\delta_n^2 f_1(\bm{p}_I^{opt}) t \!\!+\!\! f_2(\bm{p}_I^{opt})  tr( \widetilde{{\bm{H}}}_{SID} \bm{\Phi} )=1.\\
\end{array}
\end{eqnarray}
Then, in each iteration, we gradually relax the rank-one constraint and solve the corresponding convex problem.  Specifically, in the $(i+1)$-st ($i\geq0$) iteration, we solve the problem given by
\begin{eqnarray}\label{(25)}
[\bm{\Phi}^{(i+1)},t^{(i+1)}] &\!\!\!\!\!\!=\!\!\!\!\!\!&\mathrm{arg} \quad \underset{\bm{\Phi}\geq 0, t}{\mathrm{min}} g(\bm{\Phi},t)  \nonumber\\
&& {}
\begin{array}{ll}
\!\!\!\!\!\!\!\!\!\!\!\!\mathrm{s.t.} 
& \!\!\!\!\!\!0\leq \Phi_{n,n} \leq t, n=1,2,...,N\\
& \!\!\!\!\!\!\delta_n^2 f_1(\bm{p}_I^{opt}) t \!\!+\!\! f_2(\bm{p}_I^{opt})  tr( \widetilde{{\bm{H}}}_{SID} \bm{\Phi} )\!\!=\!\!1\\
&  \!\!\!\!\!\![{\bm{\mu}}_{max} (\bm{\Phi}^{(i)})]^{\rm{H}} \bm{\Phi}  {\bm{\mu}}_{max} (\bm{\Phi}^{(i)}) \!\!\geq\!\! {\omega}^{(i)} tr(\bm{\Phi} ),\\
\end{array}
\end{eqnarray}
where $0\leq{\omega}^{(i)}\leq1$ is an auxiliary parameter, which is updated in each iteration.\\
\indent Since both Problems (24) and (25) are convex, we can solve them by using the CVX toolbox. The details of the SROCR based iterative algorithm used for solving Problem (23) are presented in Table $\text{\uppercase\expandafter{\romannumeral3}}$.\\
\vspace{0.05in}
\begin{table}[!h]
\caption{  The SROCR Based Iterative Algorithm used for solving Problem (23) }
\vspace{-0.7cm}
\begin{center}
\begin{tabular}{l}
\hline
Algorithm 2: SROCR Based Iterative Algorithm.\\
\hline
\small
\textbf{Initialization}:   \\
\small
\quad  Convergence threshold $\xi_1$,  $\xi_2$, maximum iteration number $I_{iter}$.\\
\small
\quad  $i=0$, ${\omega}^{(0)}=0$. Set an initial step size  ${\delta}^{(0)}$.\\
\small
\quad Solve Problem (24) to obtain  $[\bm{\Phi}^{(0)},t^{(0)}]$. \\
\small
\textbf{Repeat} \\
\small
\quad Given  $[\bm{\Phi}^{(i)},t^{(i)},{\omega}^{(i)}]$, solve Problem (25). \\
\small
\quad \textbf{if} Problem (25) is solvable, \textbf{then} \\
\small
\quad \quad Obtain the optimal solution  $[\bm{\Phi}^{(i+1)},t^{(i+1)}]$;\\
\small
\quad \quad ${\delta}^{(i+1)} \longleftarrow {\delta}^{(0)}$ \\ 
\small
\quad  \textbf{else} \\
\small
\quad \quad ${\delta}^{(i+1)} \longleftarrow \frac{{\delta}^{(i)}}{2}$ \\ 
\small
\quad  \textbf{end if}\\
\small
\quad  ${\omega}^{(i+1)} \longleftarrow min(1,\frac{ \lambda_{max}(\bm{\Phi}^{(i+1)}) } {tr(\bm{\Phi}^{(i+1)})}+{\delta}^{(i+1)})$.\\
\small
\quad  $i \longleftarrow i+1$ \\ 
\small
\textbf{Until} ${\omega}^{(i)}\!\geq\! \xi_1 \& |g(\bm{\Phi}^{(i)},t^{(i)})\!-\!g(\bm{\Phi}^{(i-1)},t^{(i-1)}) | \!\leq\! \xi_2$ or $i \!\geq\! I_{iter}$\\
\small
\textbf{end Repeat}\\
\hline
\end{tabular}
\end{center}
\end{table}
\indent Based on the solution  $[\bm{\Phi}^{(n_{iter})},t^{(n_{iter})}]$ obtained by using the SROCR based iterative algorithm, where $n_{iter}$ denotes the number of iterations, then we can obtain the quasi-optimal solution $\bm{\Psi}^{q-opt}=\frac{\bm{\Phi}^{(n_{iter})}}{t^{(n_{iter})}}$ of Problem (23). Then, we can recover the quasi-optimal solution $\bm{q}^{q-opt}$ due to $\bm{\Psi}^{q-opt}=(\bm{q}^{q-opt})^{\rm{H}}\bm{q}^{q-opt}$.
\subsection{Discussion: Extension to a MISO system}  
In this subsection, we discuss the scenario, where the source is equipped with $ N_s$($N_s>1$) antennas, and show that the two-tier algorithm proposed in Section $\text{\uppercase\expandafter{\romannumeral3}}. A$ is eminently suitable for a MISO system. Let us denote the small-scale fading coefficient of the $S$-$I$ channel and the transmit beamforming vector by $\widetilde{{\bm{H}}}_{SI}  \in \mathbb{C}^{N\times N_s}$ and $\bm{b}_s \in \mathbb{C}^{N_s\times 1}$, respectively. Upon introducing $\widetilde{{\bm{h}}}_{SID}'=diag(\widetilde{{\bm{H}}}_{SI} \bm{b}_s) \widetilde{{\bm{h}}}_{ID}$, $\widetilde{{\bm{h}}}_{SIE}'=diag(\widetilde{{\bm{H}}}_{SI} \bm{b}_s) \widetilde{{\bm{h}}}_{IE}$, we can formulate the achievable secrecy rate of the MISO system as 
\begin{equation}\label{(25)}
C_s^{MISO}={\mathrm{max}} (0,\log_2  \frac{ \delta_n^2 f_1(\bm{p}_I ) + f_2(\bm{p}_I )  | \bm{q} {\widetilde{{\bm{h}}}_{SID}}'|^2}{\delta_n^2 f_1(\bm{p}_I) +f_3(\bm{p}_I) | \bm{q} {\widetilde{{\bm{h}}}_{SIE}}'|^2 }  ),
\end{equation}
which is similar to (7). Correspondingly, we can use the similar analysis method in Section  $\textrm{\uppercase\expandafter{\romannumeral3}}.A$ to conclude that  $d_{SI}^{\alpha_{SI}} d_{ID}^{\alpha_{ID}} $ should be minimized for the sake of maximizing the secrecy rate. Then, we can use the proposed two-tier algorithm to optimize the 3D RIS location for the MISO system.
\section{Numerical Results and Discussions}
\indent In this section, we present the secrecy rate results of the proposed J-LPB optimization scheme. The NSB-LPB  optimization scheme,  the NDB-LPB optimization scheme, as well as the R-LPB optimization scheme are used as the benchmarks. In the NSB-LPB and  NDB-LPB optimization schemes, the RIS is located near the source $S$ and the destination $D$, respectively. In the R-LPB optimization scheme, the RIS is located randomly. It is worth mentioning that  in the NSB-LPB,  NDB-LPB and R-LPB schemes, neither the $S$-$I$ link nor the $I$-$D$ link is  blocked by the obstacle, and the passive beamforming vector is optimized correspondingly for the sake of  maximizing the secrecy rate in each transmission slot. In the NSB-LPB optimization scheme, we compute the $S$-$I$ distances for all the RIS location cases, and then select the RIS location cases where the corresponding $S$-$I$ distances are the lowest thousandth as the NSB RIS location cases. The secrecy rate of the NSB-LPB optimization scheme is averaged over these NSB RIS location cases. Similarly, we compute the $I$-$D$ distances for all the RIS location cases, and select the RIS location cases where the corresponding $I$-$D$ distances are the lowest thousandth as the NDB RIS location cases. The secrecy rate of the NDB-LPB optimization scheme is averaged over these NDB RIS location cases. The secrecy rate of the R-LPB optimization scheme is averaged over 10 random location cases.   \\
\indent In our numerical simulations, $S$, $D$, $E$, the RIS and the bulky rectangular obstacle are located within a space whose length, width and height are $L_x=10$ m, $L_y=10$ m and $L_z=3$ m, respectively. The endpoint coordinates of  the space are denoted as $[0,0,0]$, $[0,0,L_z]$, $[0,L_y,0]$, $[0,L_y,L_z]$,$[L_x,0,0]$, $[L_x,0,L_z]$, $[L_x,L_y,0]$ and $[L_x,L_y,L_z]$. The four endpoint coordinates of the obstacle are $\bm{p}_o^{l,l}=[L_x/2, L_y/2,0]$, $\bm{p}_o^{u,l}=[L_x/2, L_y/2,L_z]$, $\bm{p}_o^{l,r}=[L_x/2, L_y,0]$ and $\bm{p}_o^{u,r}=[L_x/2, L_y,L_z]$. The coordinates of $S$, $D$ and $E$ are set to $\bm{p}_S=[4,6,2]$ m, $\bm{p}_D=[8,7,1.5]$ m and $\bm{p}_E=[8.5,7,1.5]$ m, respectively. We set  $\xi_0=10^{-3}$, $d_{f}=0.5$ m, $\delta_z=0.1$ m and $\delta_d=0.1$ m for the two-tier optimization problem. It is noted that $d_{f}$ is set to 0.5 m due to the fact that $d_{f}$ generally ranges from several decimeters to several meters, when $N$ is in the range of $[10, 1000]$ and the carrier frequency $f_c$ is in the range of $[1,10]$ GHz following [36][43]. The path loss factors are set as $\alpha_{SI}=\alpha_{ID}=\alpha_{IE}=2.2$. The small-scale fading coefficients $\widetilde{{\bm{h}}}_{SI}$, $\widetilde{{\bm{h}}}_{ID}$ and $\widetilde{{\bm{h}}}_{IE}$ are modeled by Rician fading coefficients, which have Rician factors of $K_{SI}=K_{ID}=K_{IE}=1$ and variances of $\delta_{SI}^2=\delta_{ID}^2=\delta_{IE}^2=1$. The noise power is set to $-40$ dBm. Furthermore, we set $x_{min}=0$,  $x_{max}=L_x$, $y_{min}=0$,  $y_{max}=L_y$,  $z_{min}=0$ and $z_{max}=L_z$ without special explanation, which means that the RIS can be positioned anywhere in the space. In this case, $d_{SI}$ is in  $[1.42, 1.55]$ m with the average value of 1.50 m in the NSB-LPB optimization scheme, and $d_{ID}$ is in $[3.61, 3.66]$ m with the average value of 3.63 m in the NDB-LPB optimization scheme. In addition, following [42], we set $\xi_1=0.999$, $\xi_2=10^{-4}$, $\delta^{(0)}=0.1$ and $I_{iter}=100$ for the SROCR based iterative algorithm.  \\
\indent Fig. 2 shows the secrecy rate versus transmit power $P_s$ of the proposed J-LPB optimization scheme, of the NSB-LPB  optimization scheme, of the NDB-LPB  optimization scheme and of the R-LPB optimization scheme for $N=16$ RIS elements. All cases in Fig. 2 clearly show that the secrecy rate of the proposed  J-LPB optimization scheme is higher than that of the NSB-LPB  optimization scheme, that of the NDB-LPB  optimization scheme and that of the R-LPB optimization scheme. In addition, observe from Fig. 2 that the secrecy rate of the NSB-LPB  optimization scheme is higher than that of the NDB-LPB  optimization scheme. The reason for this trend is that on average the NSB-LPB  optimization scheme leads to a lower $d_{SI}^{\alpha_{SI}}d_{ID}^{\alpha_{ID}}$ than the NDB-LPB  optimization scheme  in our numerical simulation scenario. Moreover, from Fig. 2, we see that as the transmit power $P_s$ increases, the secrecy rate of all the four schemes improves. 
\\    
\begin{figure}[htb!]
\centering
\includegraphics[scale=0.65]{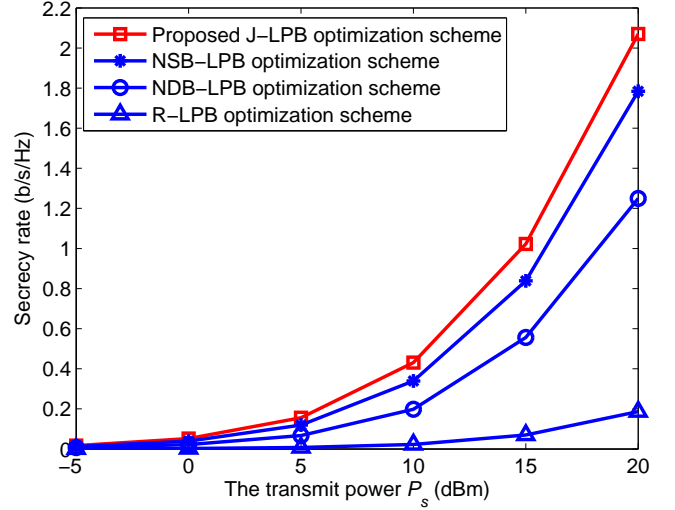}
\caption{ Secrecy rate versus transmit power $P_s$  of the proposed J-LPB optimization scheme, the NSB-LPB  optimization scheme, the NDB-LPB  optimization scheme and the R-LPB optimization scheme for $N=16$.    }
\label{fig_2}
\end{figure}
\indent In Fig. 3, we depict the secrecy rate versus the number of RIS elements $N$ of the proposed J-LPB optimization scheme, of the NSB-LPB  optimization scheme, of the NDB-LPB  optimization scheme and of the R-LPB optimization scheme for $P_s=10$ dBm.  Observe from Fig. 3 that the proposed J-LPB optimization scheme outperforms the NSB-LPB  optimization scheme, the NDB-LPB  optimization scheme and the R-LPB optimization scheme in terms of its secrecy rate. Fig. 3 also shows that as $N$ increases, the advantage of the proposed J-LPB optimization scheme over both the NSB-LPB  optimization scheme and over the NDB-LPB optimization scheme becomes more obvious, which illustrates that the superiority of our proposed J-LPB optimization scheme can be further improved by increasing the number of RIS elements.  \\
\begin{figure}[htb!]
\centering
\includegraphics[scale=0.65]{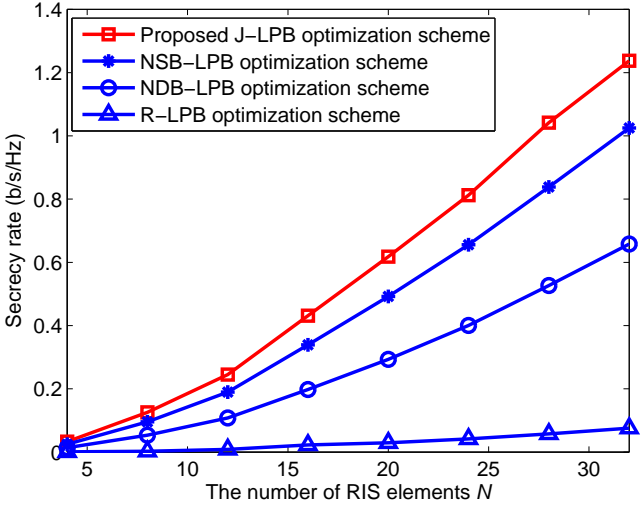}
\caption{ Secrecy rate versus the number of RIS elements $N$ of  the proposed J-LPB optimization scheme, the NSB-LPB  optimization scheme, the NDB-LPB  optimization scheme and the R-LPB optimization scheme for $P_s=10$ dBm.   }
\label{fig_3}
\end{figure}
\indent Fig. 4 shows the secrecy rate versus the maximum value of the $x$-coordinate of the RIS $x_{max}$  for the proposed J-LPB optimization scheme, for the NSB-LPB  optimization scheme, for the NDB-LPB  optimization scheme and for the R-LPB optimization scheme for $P_s=10$ dBm and $N=16$ RIS elements.  The minimum value of the $x$-coordinate is set to 0. It is observed that the proposed J-LPB optimization scheme achieves a higher secrecy rate than the other three benchmark schemes, explicitly showing the advantage of the joint optimization of the RIS location and passive beamforming.  We can also observe from Fig. 4 that as the maximum value of the $x$-coordinate of the RIS increases, the secrecy rate of the proposed  J-LPB optimization scheme first improves and then saturates. This is because upon increasing $x_{max}$, the space in which the RIS can be deployed is larger, hence a better RIS location leading to a higher secrecy rate  can be found.  On the other hand, the optimal 3D RIS coordinate is $\bm{p}_I^{opt}=[4.88,4.92,2.10]$ m, when the RIS can be deployed anywhere in the 10m$\times$10m$\times$3m space in our simulation set, which means that the optimal RIS location can be achieved as long as $x_{max}$ is higher than 4.88 m. It is worth noting that the secrecy rate of the R-LPB optimization scheme fluctuates since the random RIS locations are different in different $x_{max}$ cases.   \\
\begin{figure}[htb!]
\centering
\includegraphics[scale=0.65]{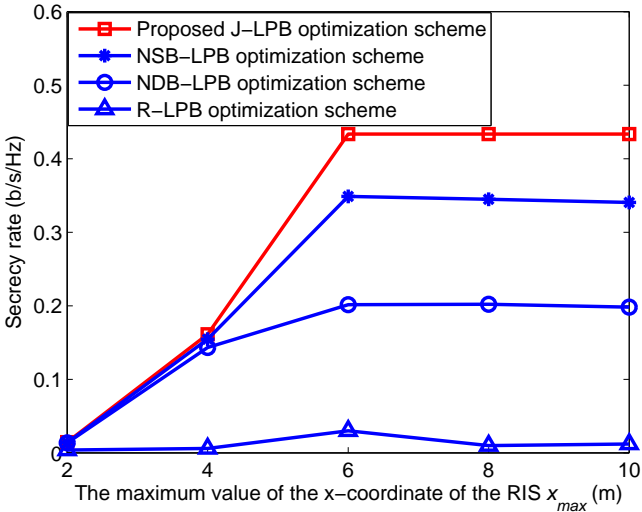}
\caption{ Secrecy rate versus the maximum value of the x-coordinate of the RIS $x_{max}$ for the proposed J-LPB optimization scheme, the NSB-LPB  optimization scheme, the NDB-LPB  optimization scheme and the R-LPB optimization scheme for $P_s=10$ dBm and $N=16$.  }
\label{fig_4}
\end{figure}
\indent In Fig. 5, we show the secrecy rate versus the maximum value of the $y$-coordinate of the RIS $y_{max}$  for the proposed J-LPB optimization scheme, for the NSB-LPB  optimization scheme, for the NDB-LPB  optimization scheme and for the R-LPB optimization scheme for $P_s=10$ dBm and $N=16$ reflecting units at the RIS. One can see from Fig. 5 that the proposed J-LPB optimization scheme outperforms all the other three benchmark schemes in terms of its secrecy rate. Additionally, Fig. 5 also shows that the secrecy rate of the proposed  J-LPB optimization scheme first increases with $y_{max}$ and then saturates. The reason for this is that a larger $y_{max}$ leads to a larger space in which the RIS can be deployed to achieve a higher secrecy rate. On the other hand, since the optimal RIS coordinate is $\bm{p}_I^{opt}=[4.88,4.92,2.10]$ m in the scenario, where the RIS can be deployed anywhere in the 10m$\times$10m$\times$3m space, the optimal RIS location $\bm{p}_I^{opt}=[4.88,4.92,2.10]$ m can always be obtained as long as $y_{max}\geq 4.92$ m is satisfied. \\  
\begin{figure}[htb!]
\centering
\includegraphics[scale=0.65]{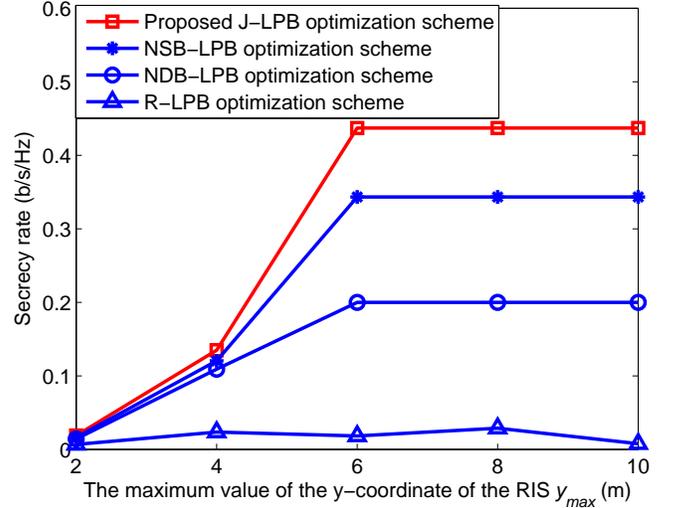}
\caption{ Secrecy rate versus the maximum value of the y-coordinate of the RIS $y_{max}$ for the proposed J-LPB optimization scheme, the NSB-LPB  optimization scheme, the NDB-LPB  optimization scheme and the R-LPB optimization scheme for $P_s=10$ dBm and $N=16$.  }
\label{fig_5}
\end{figure}
\indent Fig. 6 illustrates the secrecy rate versus the maximum value of the $z$-coordinate of the RIS $z_{max}$ for the proposed J-LPB optimization scheme, for the NSB-LPB  optimization scheme, for the NDB-LPB  optimization scheme and for the R-LPB optimization scheme for $P_s=10$ dBm and $N=16$ RIS units. It can be observed from Fig. 6 that the proposed J-LPB optimization scheme always performs better than the NSB-LPB optimization scheme, the NDB-LPB  optimization scheme and the R-LPB optimization scheme. Additionally, as shown in Fig. 6, the secrecy rate of the proposed J-LPB optimization scheme increases with $z_{max}$ when $z_{max} \leq 2$m since a higher $z_{max}$ implies a larger room where the RIS can be better positioned to achieve a higher secrecy rate. We can also see from Fig. 6 that when $z_{max} \geq 2.0$ m, the secrecy rate of the proposed J-LPB optimization scheme saturates near constant. This is because the optimal RIS location  $\bm{p}_I^{opt}=[4.88,4.92,2.10]$ m in the scenario, where the RIS can be deployed anywhere in the 10m$\times$10m$\times$3m space, can be always obtained as long as $z_{max}\geq 2.10$ m, which is close to 2.0 m.    
\begin{figure}[htb!]
\centering
\includegraphics[scale=0.65]{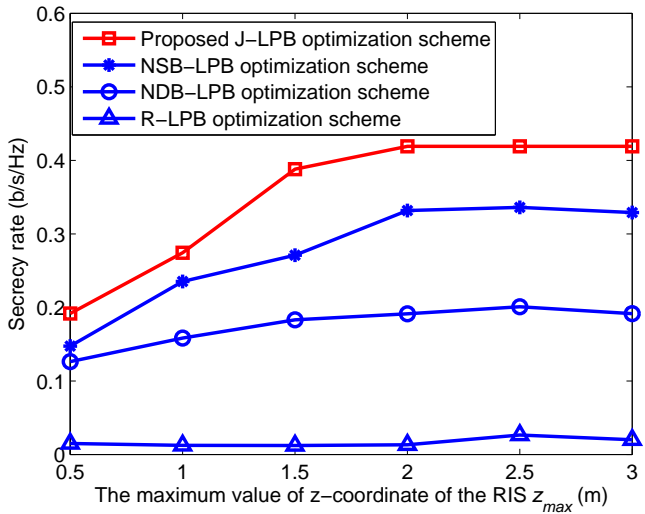}
\caption{ Secrecy rate versus the maximum value of the z-coordinate of the RIS $z_{max}$ for the proposed J-LPB optimization scheme, the NSB-LPB  optimization scheme, the NDB-LPB  optimization scheme and the R-LPB optimization scheme for $P_s=10$ dBm and $N=16$.  }
\label{fig_6}
\end{figure}  
\section{Conclusions}
\indent The PLS of wireless networks was investigated where a small RIS was deployed to help a source to send confidential information to a legitimate user in the presence of a malicious eavesdropper. We proposed a J-LPB optimization scheme, where the RIS location and passive beamforming are jointly optimized for maximizing the system's secrecy rate under the RIS location constraint and the constraint that the modulus of the passive beamforming coefficient at each RIS's unit is not larger than 1.  We analyzed the optimal RIS location, and concluded that the 3D RIS coordinate should be optimized for the sake of minimizing the product involving the source-RIS distance and the RIS-destination distance. Then, we proposed a two-tier optimization algorithm to solve the associated non-convex product minimization problem to obtain the near-optimal 3D RIS coordinate. On this basis, we further formulated the passive beamforming optimization problem, and proposed to solve it by using the Charnes-Cooper transformation together with the SROCR algorithm. Our simulation results demonstrate that the proposed J-LPB optimization scheme achieves a higher average secrecy rate than the NSB-LPB optimization scheme, the NDB-LPB optimization scheme and the R-LPB optimization scheme. Additionally, our simulation results show that the  secrecy rate of the J-LPB optimization scheme generally increases with  the number of reflecting units and the deployment space.   
\appendix
\indent Upon introducing $\bm{B}_1={\widetilde{{\bm{h}}}_{SID}} {\widetilde{{\bm{h}}}_{SID}}^{\rm{H}}$ and $\bm{B}_2=-\frac{ d_{ID}^{\alpha_{ID}} } {d_{IE}^{\alpha_{IE}} } {\widetilde{{\bm{h}}}_{SIE}} {\widetilde{{\bm{h}}}_{SIE}}^{\rm{H}}$, we have $\bm{B}=\bm{B}_1+\bm{B}_2$. Let us now prove that $\bm{B}$ is NOT a negative semidefinite matrix.\\
\indent Since $rank(\bm{B}_1)=1$, $rank(\bm{B}_2)=1$ and $rank(\bm{B})\leq rank(\bm{B}_1)+rank(\bm{B}_2)$, we have $rank(\bm{B})\leq 2$. Thus, we can readily see that the square matrix $\bm{B}$ has at most two non-zero eigenvalues. \\
\indent Let us sort the eigenvalues of $\bm{B}_1$ as well as those of $\bm{B}_2$ in ascending order, and denote them as $\lambda_1(\bm{B}_1) \leq \lambda_2(\bm{B}_1) \leq ... \leq \lambda_N(\bm{B}_1)$ and $\lambda_1(\bm{B}_2) \leq \lambda_2(\bm{B}_2) \leq ... \leq \lambda_N(\bm{B}_2)$, respectively. Since $rank(\bm{B}_1)=1$ and $tr(\bm{B}_1)=|{\widetilde{{\bm{h}}}_{SID}}|^2$, we can see that $\bm{B}_1$ has only a single non-zero eigenvalue $|{\widetilde{{\bm{h}}_{SID}}}|^2$. That is,
\renewcommand\theequation{A.1}
\begin{equation}
\lambda_n(\bm{B}_1) =
\left\{
\begin{array} {lr}
0,&1 \leq n\leq N-1 \\
|{\widetilde{{\bm{h}}}_{SID}}|^2, &n=N
\end{array}. 
\right.
\end{equation}
Similarly, we  have
\renewcommand\theequation{A.2}
\begin{equation}
\lambda_n(\bm{B}_2) =
\left\{
\begin{array} {lr}
-\frac{ d_{ID}^{\alpha_{ID}} } {d_{IE}^{\alpha_{IE}} } |{\widetilde{{\bm{h}}}_{SIE}}|^2,&n=1 \\
0, &2\leq n \leq N
\end{array}. 
\right.
\end{equation}
\indent Let us sort the eigenvalues of $\bm{B}$ in ascending order, and represent them as $\lambda_1(\bm{B}) \leq \lambda_2(\bm{B}) \leq ... \leq \lambda_N(\bm{B})$. Since both $\bm{B}_1$ and $\bm{B}_2$ are Hermitian matrices, following the Weyl Formula [44], we arrive at  
\renewcommand\theequation{A.3}
\begin{equation}
\lambda_n(\bm{B}) \geq 
\left\{
\begin{array} {l}
\lambda_n(\bm{B}_1)+ \lambda_1(\bm{B}_2) \\
\lambda_{n-1}(\bm{B}_1)+ \lambda_2(\bm{B}_2) \\
\vdots  \\
\lambda_1(\bm{B}_1)+ \lambda_n(\bm{B}_2) 
\end{array} 
\right.
\end{equation}
and\\
\renewcommand\theequation{A.4}
\begin{equation}
\lambda_n(\bm{B}) \leq 
\left\{
\begin{array} {l}
\lambda_n(\bm{B}_1)+ \lambda_N(\bm{B}_2) \\
\lambda_{n+1}(\bm{B}_1)+ \lambda_{N-1}(\bm{B}_2) \\
\vdots  \\
\lambda_{N}(\bm{B}_1)+ \lambda_n(\bm{B}_2) 
\end{array}, 
\right.
\end{equation}
where $n\in [1, N]$.\\
\indent By substituting (A.1) and (A.2) into (A.3) and (A.4), we arrive at (A.5) stated on the following page.
\begin{figure*}
\renewcommand\theequation{A.5}
\begin{equation}
\lambda_n(\bm{B}) \in 
\left\{
\begin{array} {cr}
\left[ -\frac{ d_{ID}^{\alpha_{ID}} } {d_{IE}^{\alpha_{IE}} } |{\widetilde{{\bm{h}}}_{SIE}}|^2, min(0, |{\widetilde{{\bm{h}}}_{SID}}|^2-\frac{ d_{ID}^{\alpha_{ID}} } {d_{IE}^{\alpha_{IE}} } |{\widetilde{{\bm{h}}}_{SIE}}|^2) \right], &n=1 \\
\left[0,  |{\widetilde{{\bm{h}}}_{SID}}|^2 \right],&2\leq n \leq N-1 \\
\left[ max(0, |{\widetilde{{\bm{h}}}_{SID}}|^2-\frac{ d_{ID}^{\alpha_{ID}} } {d_{IE}^{\alpha_{IE}} } |{\widetilde{{\bm{h}}}_{SIE}}|^2), |{\widetilde{{\bm{h}}}_{SID}}|^2 \right], & n=N
\end{array} 
\right.
\end{equation}
\noindent\rule[-10pt]{\textwidth}{0.5pt}
\end{figure*}\\
\indent According to the value of  $|{\widetilde{{\bm{h}}}_{SID}}|^2-\frac{ d_{ID}^{\alpha_{ID}} } {d_{IE}^{\alpha_{IE}} } |{\widetilde{{\bm{h}}}_{SIE}}|^2$, we discuss the eigenvalues of $\bm{B}$ in three cases as follows.
\begin{enumerate} 
\item Let us first discuss the case where $|{\widetilde{{\bm{h}}}_{SID}}|^2-\frac{ d_{ID}^{\alpha_{ID}} } {d_{IE}^{\alpha_{IE}} } |{\widetilde{{\bm{h}}}_{SIE}}|^2>0$. In this case,  we have $max(0, |{\widetilde{{\bm{h}}}_{SID}}|^2-\frac{ d_{ID}^{\alpha_{ID}} } {d_{IE}^{\alpha_{IE}} } |{\widetilde{{\bm{h}}}_{SIE}}|^2)=|{\widetilde{{\bm{h}}}_{SID}}|^2-\frac{ d_{ID}^{\alpha_{ID}} } {d_{IE}^{\alpha_{IE}} } |{\widetilde{{\bm{h}}}_{SIE}}|^2>0$. Thus, from (A. 5), we can obtain that $\lambda_N(\bm{B})>0$. That is, the matrix $\bm{B}$ has at least one positive eigenvalue. Thus, $\bm{B}$ is NOT a negative semidefinite matrix in this case.
\item  Let us now discuss the case where $|{\widetilde{{\bm{h}}}_{SID}}|^2-\frac{ d_{ID}^{\alpha_{ID}} } {d_{IE}^{\alpha_{IE}} } |{\widetilde{{\bm{h}}}_{SIE}}|^2=0$. In this case, we have 
\renewcommand\theequation{A.6}
\begin{equation}
{\sum_{n=1}^N} {\lambda_n(\bm{B})}=tr(\bm{B})=|{\widetilde{{\bm{h}}}_{SID}}|^2-\frac{ d_{ID}^{\alpha_{ID}} } {d_{IE}^{\alpha_{IE}} } |{\widetilde{{\bm{h}}}_{SIE}}|^2=0.
\end{equation}
\indent According to (A. 5), we have $\lambda_N(\bm{B})\geq 0$. If $\lambda_N(\bm{B})> 0$, we can obviously obtain that $\bm{B}$ is not a negative semidefinite matrix since it has a positive eigenvalue. If $\lambda_N(\bm{B})= 0$, by combining $\lambda_1(\bm{B}) \leq \lambda_2(\bm{B}) \leq ... \leq \lambda_N(\bm{B})$ and (A. 6), we can readily obtain that $\lambda_n(\bm{B})= 0$ ($1 \leq n \leq N$). That is, $\bm{B}$ is a zero matrix. \\
\indent Thus, $\bm{B}$ is NOT a negative semidefinite matrix in this case.
\item Let us finally discuss the case where $|{\widetilde{{\bm{h}}}_{SID}}|^2-\frac{ d_{ID}^{\alpha_{ID}} } {d_{IE}^{\alpha_{IE}} } |{\widetilde{{\bm{h}}}_{SIE}}|^2<0$. In this case, (A. 5) can be rewitten as (A. 7) stated on the following page.
\begin{figure*}
\renewcommand\theequation{A.7}
\begin{equation}
\lambda_n(\bm{B}) \in 
\left\{
\begin{array} {cr}
\left[ -\frac{ d_{ID}^{\alpha_{ID}} } {d_{IE}^{\alpha_{IE}} } |{\widetilde{{\bm{h}}}_{SIE}}|^2,  |{\widetilde{{\bm{h}}}_{SID}}|^2-\frac{ d_{ID}^{\alpha_{ID}} } {d_{IE}^{\alpha_{IE}} } |{\widetilde{{\bm{h}}}_{SIE}}|^2\right], &n=1 \\
\left[0,  |{\widetilde{{\bm{h}}}_{SID}}|^2 \right],&2\leq n \leq N 
\end{array} 
\right.
\end{equation}
\noindent\rule[-10pt]{\textwidth}{0.5pt}
\end{figure*}
Moreover, we have
\renewcommand\theequation{A.8}
\begin{equation}
{\sum_{n=1}^N} {\lambda_n(\bm{B})}=|{\widetilde{{\bm{h}}}_{SID}}|^2-\frac{ d_{ID}^{\alpha_{ID}} } {d_{IE}^{\alpha_{IE}} } |{\widetilde{{\bm{h}}}_{SIE}}|^2<0.
\end{equation}
\indent \quad Since $\bm{B}$ has at most two non-zero eigenvalues and $ \lambda_1(\bm{B})$ is the smaller eigenvalue of $\bm{B}$, by combining (A.7) and (A.8), we obtain that  $\lambda_n(\bm{B})$ ($1 \leq n \leq N$) can only take one of two value sets, which are $ \lambda_1(\bm{B})<|{\widetilde{{\bm{h}}}_{SID}}|^2-\frac{ d_{ID}^{\alpha_{ID}} } {d_{IE}^{\alpha_{IE}} } |{\widetilde{{\bm{h}}}_{SIE}}|^2$, $ \lambda_n(\bm{B})=0$ ($2 \leq n \leq N-1$), $ \lambda_N(\bm{B})>0$ and $\lambda_1(\bm{B})=|{\widetilde{{\bm{h}}}_{SID}}|^2-\frac{ d_{ID}^{\alpha_{ID}} } {d_{IE}^{\alpha_{IE}} } |{\widetilde{{\bm{h}}}_{SIE}}|^2$,  $ \lambda_n(\bm{B})=0$ ($2 \leq n \leq N$). When $\lambda_n(\bm{B})$ takes the first set of values, we can readily see that $\bm{B}$ is not a negative semidefinite matrix due to $\lambda_N(\bm{B})>0$. \\
\indent \quad Now let us prove that $\lambda_n(\bm{B})$ cannot take the second set of values, which are $\lambda_1(\bm{B})=|{\widetilde{{\bm{h}}}_{SID}}|^2-\frac{ d_{ID}^{\alpha_{ID}} } {d_{IE}^{\alpha_{IE}} } |{\widetilde{{\bm{h}}}_{SIE}}|^2$,  $ \lambda_n(\bm{B})=0$ ($2 \leq n \leq N$). That is, $ |{\widetilde{{\bm{h}}}_{SID}}|^2-\frac{ d_{ID}^{\alpha_{ID}} } {d_{IE}^{\alpha_{IE}} } |{\widetilde{{\bm{h}}}_{SIE}}|^2$ cannot be an eigenvalue of $\bm{B}$. The proof details are given as follows.\\
\indent \quad Let us define $\bm{x} \in \mathcal{C}^{N \times 1}$ as a vector, which satisfies
\renewcommand\theequation{A.9}
\begin{equation}
\left[ 
\begin{array}{c}
	{\widetilde{{\bm{h}}}_{SID}}^{\rm{H}}\\
	{\widetilde{{\bm{h}}}_{SIE}}^{\rm{H}}
\end{array} 
\right] \bm{x}=\bm{0}_{2\times 1},
\end{equation}
where $\bm{0}_{m_1\times 1}$ denotes a zero vector of  dimension $m_1\times 1$. Let us introduce $a=\frac{ d_{ID}^{\alpha_{ID}} } {d_{IE}^{\alpha_{IE}} }$ for convenience. Then, we have 
\renewcommand\theequation{A.10}
\begin{align}
\bm{B} \bm{x}&={\widetilde{{\bm{h}}}_{SID}}{\widetilde{{\bm{h}}}_{SID}}^{\rm{H}}\bm{x}-a{\widetilde{{\bm{h}}}_{SIE}}{\widetilde{{\bm{h}}}_{SIE}}^{\rm{H}}\bm{x}   \nonumber \\
&=0 \cdot {\widetilde{{\bm{h}}}_{SID}}-0 \cdot a{\widetilde{{\bm{h}}}_{SIE}}=\bm{0}_{N\times 1}.
\end{align}
From (A.10), we can see that the vector $\bm{x}$ which satisfies (A.9) is the eigenvector of $\bm{B}$, and its corresponding eigenvalue is 0. As $\bm{B}$ has at most $N-2$ zero eigenvalues, from (A. 9) and (A. 10), we can see that these $N-2$ eigenvectors whose corresponding eigenvalues are 0  are orthogonal to both ${\widetilde{{\bm{h}}}_{SID}}$ and ${\widetilde{{\bm{h}}}_{SIE}}$. Thus, since $\bm{B}$ is a Hermitian matrix, the eigenvector $\bm{e}$ of $\bm{B}$, whose corresponding eigenvalue is non-zero, must be in the space spanned by ${\widetilde{{\bm{h}}}_{SID}}$ and ${\widetilde{{\bm{h}}}_{SIE}}$.\\
\indent \quad Let us now define $\bm{e}={\widetilde{{\bm{h}}}_{SID}}+b{\widetilde{{\bm{h}}}_{SIE}}$ or $\bm{e}={\widetilde{{\bm{h}}}_{SIE}}+b{\widetilde{{\bm{h}}}_{SID}}$, where $b$ is a scalar. Here we take $\bm{e}={\widetilde{{\bm{h}}}_{SID}}+b{\widetilde{{\bm{h}}}_{SIE}}$ for example, and derive its corresponding eigenvalue $\lambda_e$ as follows. \\
\indent \quad Upon substituting  $\bm{B}={\widetilde{{\bm{h}}}_{SID}} {\widetilde{{\bm{h}}}_{SID}}^{\rm{H}}-a{\widetilde{{\bm{h}}}_{SIE}} {\widetilde{{\bm{h}}}_{SIE}}^{\rm{H}}$ and $\bm{e}={\widetilde{{\bm{h}}}_{SID}}+b{\widetilde{{\bm{h}}}_{SIE}}$ into $\bm{B} \bm{e}=\lambda_e \bm{e}$, we have
\renewcommand\theequation{A.11}
\begin{align}
&( {\widetilde{{\bm{h}}}_{SID}} {\widetilde{{\bm{h}}}_{SID}}^{\rm{H}}-a{\widetilde{{\bm{h}}}_{SIE}}  {\widetilde{{\bm{h}}}_{SIE}}^{\rm{H}})  ({\widetilde{{\bm{h}}}_{SID}}+b{\widetilde{{\bm{h}}}_{SIE}})  \nonumber \\
&={\widetilde{{\bm{h}}}_{SID}}^{\rm{H}} {\widetilde{{\bm{h}}}_{SID}} {\widetilde{{\bm{h}}}_{SID}}-a{\widetilde{{\bm{h}}}_{SIE}}^{\rm{H}} {\widetilde{{\bm{h}}}_{SID}} {\widetilde{{\bm{h}}}_{SIE}}  \nonumber \\
&+b{\widetilde{{\bm{h}}}_{SID}}^{\rm{H}}{\widetilde{{\bm{h}}}_{SIE}}{\widetilde{{\bm{h}}}_{SID}}-ab{\widetilde{{\bm{h}}}_{SIE}}^{\rm{H}}{\widetilde{{\bm{h}}}_{SIE}}{\widetilde{{\bm{h}}}_{SIE}} \nonumber \\
&=\lambda_e ({\widetilde{{\bm{h}}}_{SID}}+b{\widetilde{{\bm{h}}}_{SIE}}).
\end{align}
\indent \quad Upon denoting $c_1={\widetilde{{\bm{h}}}_{SID}}^{\rm{H}} {\widetilde{{\bm{h}}}_{SID}}\geq 0$, $c_2={\widetilde{{\bm{h}}}_{SIE}}^{\rm{H}} {\widetilde{{\bm{h}}}_{SID}}$ and $c_3={\widetilde{{\bm{h}}}_{SIE}}^{\rm{H}} {\widetilde{{\bm{h}}}_{SIE}} \geq 0$, we can rewrite (A.11) as
\renewcommand\theequation{A.12}
\begin{equation}
(c_1+bc_2^{\rm{H}}){\widetilde{{\bm{h}}}_{SID}}-(ac_2+abc_3){\widetilde{{\bm{h}}}_{SIE}}=\lambda_e {\widetilde{{\bm{h}}}_{SID}} +b\lambda_e {\widetilde{{\bm{h}}}_{SIE}}.
\end{equation}
From (A.12), we can readily infer that
\renewcommand\theequation{A.13}
\begin{equation}
\lambda_e=c_1+bc_2^{\rm{H}}
\end{equation}
and
\renewcommand\theequation{A.14}
\begin{equation}
-b\lambda_e=ac_2+abc_3,
\end{equation}
where $b$ and $\lambda_e$ are the unknown variables. When $c_2 \neq 0$, by solving (A.13) and (A.14), we have
\renewcommand\theequation{A.15}
\begin{equation}
\lambda_e=\frac{c_1-ac_3}{2} \pm \frac{|c_2|}{2} \sqrt{\frac{(c_1+ac_3)^2}{|c_2|^2}-4a }.
\end{equation}
Since ${\widetilde{{\bm{h}}}_{ID}}$ is not completely correlated with ${\widetilde{{\bm{h}}}_{IE}}$, we have ${\widetilde{{\bm{h}}}_{ID}} \neq l{\widetilde{{\bm{h}}}_{IE}}$ ($\forall l \in \mathcal{R}$). Thus, due to $\widetilde{{\bm{h}}}_{SID}=diag(\widetilde{{\bm{h}}}_{SI}) \widetilde{{\bm{h}}}_{ID}$ and $\widetilde{{\bm{h}}}_{SIE}=diag(\widetilde{{\bm{h}}}_{SI}) \widetilde{{\bm{h}}}_{IE}$,  we can have ${\widetilde{{\bm{h}}}_{SID}} \neq l {\widetilde{{\bm{h}}}_{SIE}}$. Then, we can obtain that $c_1c_3 \neq |c_2|^2$, which leads to $(c_1-ac_3)^2 \neq |c_2|^2 \lbrack \frac{(c_1+ac_3)^2}{|c_2|^2}-4a \rbrack$. Hence we can further arrive at
\renewcommand\theequation{A.16}
\begin{equation}
\lambda_e \neq c_1-ac_3.
\end{equation}
\indent \quad In a similar way, we can also show that (A.16) still holds when  $\bm{e}={\widetilde{{\bm{h}}}_{SIE}}+b{\widetilde{{\bm{h}}}_{SID}}$.  
That is, $ |{\widetilde{{\bm{h}}}_{SID}}|^2-\frac{ d_{ID}^{\alpha_{ID}} } {d_{IE}^{\alpha_{IE}} } |{\widetilde{{\bm{h}}}_{SIE}}|^2$ cannot be an eigenvalue of $\bm{B}$.\\
\indent Thus, in the case where $|{\widetilde{{\bm{h}}}_{SID}}|^2-\frac{ d_{ID}^{\alpha_{ID}} } {d_{IE}^{\alpha_{IE}} } |{\widetilde{{\bm{h}}}_{SIE}}|^2<0$, $\lambda_n(\bm{B})$ ($1 \leq n \leq N$) can only take the first set of values, which are $ \lambda_1(\bm{B})<|{\widetilde{{\bm{h}}}_{SID}}|^2-\frac{ d_{ID}^{\alpha_{ID}} } {d_{IE}^{\alpha_{IE}} } |{\widetilde{{\bm{h}}}_{SIE}}|^2$, $ \lambda_n(\bm{B})=0$ ($2 \leq n \leq N-1$) and $ \lambda_N(\bm{B})>0$. Thus, we can see that $\bm{B}$ is NOT a negative semidefinite matrix in this case, since $ \lambda_N(\bm{B})$ is positive.\\
\end{enumerate} 
\indent \quad \quad As discussed above, we can infer that $\bm{B}$ is NOT a negative semidefinite matrix.

\end{document}